\documentstyle[nato]{crckapb}      
\input epsf

\def\period      {{\rm ~.}}
\def\comma      {{\rm ~,}}

\def\lsim   {_\sim^<}    
\def\gsim   {_\sim^>}    
\def\etal   {{\rm et~al.}}
\def\aj     {Ap. J.}
\def\araa   {Ann. Rev. Astron. \& Astrophysics}
\def\aap    {Astronmy \& Astrophysics}
\def\pr     {Phy. Rev.}
\def\prl    {Phy. Rev. Let.}
\def\ppnp   {ppnp}

\begin{opening}
\title{THE CMB SPECTRUM}
\subtitle{Cosmic Microwave Background}

\author{George F. Smoot}
\institute{Lawrence Berkeley National Lab \& Physics Department\\
           University of California \\ Berkeley CA 94720}

\end{opening}

\runningtitle{THE CMB SPECTRUM}

\begin{document}


\section{Introduction}
The observed cosmic microwave background (CMB) radiation provides
strong evidence for the hot big bang.  The success of
primordial nucleosynthesis calculations
(``Big-bang nucleosynthesis'')
requires a cosmic background radiation (CBR) characterized
by a temperature $kT \sim 1\,$MeV at a redshift of $z \simeq 10^9$.
In their pioneering work, Gamow, Alpher, and Herman\cite{Gamow48} realized
this and predicted the existence of a faint residual relic of the
primordial radiation, with a present temperature of a few degrees.
The observed CMB is interpreted as the current manifestation
of the hypothesized CBR.

The CMB was serendipitously discovered by Penzias and Wilson\cite{Penzias65}
in 1964.
Its spectrum is well characterized by a $2.73 \pm 0.01\,$K black-body
(Planckian) spectrum over more than three decades in frequency
(see Figure 1) 
A non-interacting Planckian distribution of temperature $T_i$ at redshift $z_i$ transforms
with the universal expansion
to another Planckian distribution at redshift $z_r$ with temperature
$T_r / (1+ z_r)= T_i / (1+ z_i)$.
Hence thermal equilibrium, once established (e.g.~at the nucleosynthesis
epoch), is preserved by the expansion, in spite of the fact that photons
decoupled from matter at early times.
Because there are about $10^9$ photons per nucleon,
the transition from the ionized primordial plasma to neutral atoms at
$z\sim1000$ does not
significantly alter the CBR spectrum\cite{Peebles93}.
\begin{figure}
\centerline{\epsfxsize= 9 cm \epsfbox[148 230 479 471]{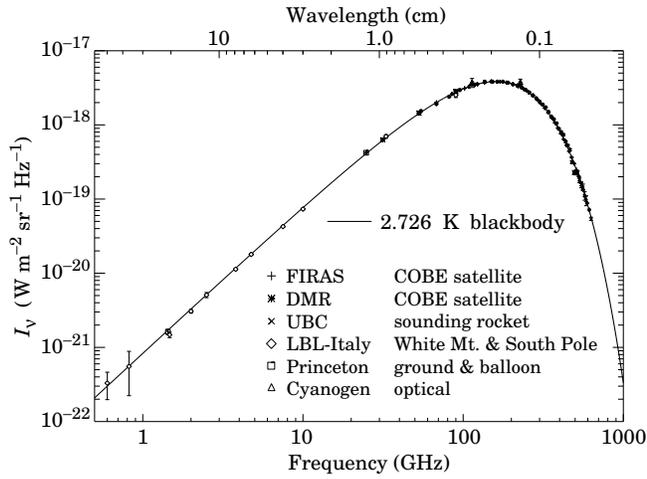}}
\caption{Precise measurements of the CMB spectrum.
The line represents a 2.73~K blackbody,
which describes the spectrum very well,
especially around the peak of intensity.
The spectrum is less well constrained
at frequencies of 3 GHz and below 
($10\,$cm and longer wavelengths).
(References for this figure are at the end of this section under
``CMB Spectrum References.'')
}
\end{figure}

\begin{figure}
\centerline{\epsfxsize=9 cm\epsfbox[115 72 422 320]{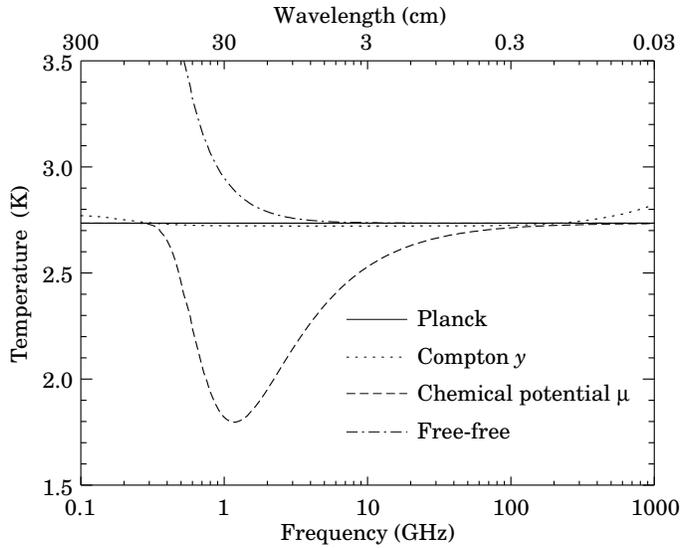}}
\caption{The shapes of expected, but so far unobserved,
CMB distortions, resulting from energy-releasing processes
at different epochs.}
\end{figure}

\section{Theoretical spectral distortions}
The remarkable precision with which the CMB spectrum is fitted by a Planckian
distribution
provides limits on possible energy releases in the early Universe,
at roughly the fractional level of $10^{-4}$ of the CBR energy,
for redshifts $\lsim 10^7$ (corresponding to epochs $\gsim 1\,$year).
The following three important classes of spectral distortions (see
Figure 2 
generally correspond to energy releases at different epochs.
The distortion results from interactions with a hot electron gas
at temperature $T_e$.

\subsection{Compton distortion} {\it Late energy release ($z\sim10^5$).}
Compton scattering
($\gamma e \rightarrow \gamma' e'$)
of the CBR photons by a hot electron gas creates spectral distortions
by transferring energy from the electrons to the photons.
Compton scattering cannot achieve thermal equilibrium
for $y < 1$, where
\begin{equation}
y = \int^z_0 ~\frac{ kT_e(z')
- kT_{\gamma}(z') }{ m_e c^2}\; \sigma_T\; n_e(z')
\; c\; \frac{dt}{dz'} \;dz'\comma
\end{equation}
is the integral of the number of interactions, $\sigma_T\; n_e(z)\; c\; dt$,
times the mean-fractional photon-energy change per collision\cite{sunzel80}.
For $T_e\gg T_\gamma$ $y$ is also proportional to the integral of the
electron pressure $n_e k T_e$ along the line of sight.
For standard thermal histories
$y < 1$ for epochs later than $z\simeq10^5$.

The resulting CMB distortion is approximately a temperature decrement
\begin{equation}
\Delta T_{\rm RJ} = -2y\;T_\gamma
\end{equation}
in the Rayleigh-Jeans ($h\nu/kT_\gamma \ll 1$) portion of the spectrum,
and a rapid rise in temperature in
the Wien ($h\nu/kT_\gamma \gg 1$) region, i.e.~photons are shifted from low to
high frequencies.
The magnitude of the distortion is related
to the total energy transfer\cite{sunzel80}
$\Delta E$ by
\begin{equation}
\Delta E/E_{\rm CBR} = e^{4y} - 1 \simeq 4y\period
\end{equation}
A prime candidate for producing a Comptonized spectrum is a hot
intergalactic medium.
A hot ($T_e>10^5\,$K) medium in clusters of galaxies also produces a
partially Comptonized spectrum as seen through the cluster, known as the
Sunyaev-Zel'dovich effect.
Based upon X-ray data, the predicted large angular scale total combined effect
of the hot intracluster medium should produce $y < 10^{-6}$\cite{Ceb94}.

\subsection{Bose-Einstein or chemical potential distortion}
{\it Early energy release ($z\sim10^5$--$10^7$).}
After many Compton scatterings ($y > 1$), the photons and electrons will reach
statistical (not thermodynamic) equilibrium,
because Compton scattering conserves photon number.  This equilibrium is
described by the Bose-Einstein distribution with non-zero chemical
potential:
\begin{equation}
n = \frac{1}{ e^{x + \mu_0} - 1 }\comma
\end{equation}
where  $x \equiv { h \nu / k T}$  and
$\mu_0 \simeq 1.4 {\Delta E} / E_{\rm CBR}$, with
$\mu_0$ being the dimensionless chemical potential that
is required to conserve photon number.

The collisions of electrons with nuclei in the plasma produce
free-free (thermal bremsstrahlung) radiation:
$ e Z \rightarrow e   Z  \gamma $.
Free-free emission thermalizes the spectrum to the plasma temperature
at long wavelengths.
Including this effect, the chemical potential becomes frequency-dependent,
\begin{equation}
\mu(x) = \mu_0 e^{- 2x_b/x }\comma
\end{equation}
where $x_b$ is the dimensionless transition frequency at which
Compton scattering of photons to higher frequencies
is balanced by free-free creation of new photons.
The resulting spectrum has a sharp drop in brightness temperature
at centimeter wavelengths\cite{burigana91}.
The minimum wavelength is determined by $\Omega_B$.

The equilibrium
Bose-Einstein distribution results from the oldest non-equilibrium
processes $(10^5 < z < 10^7)$,
such as the decay of relic particles or primordial inhomogeneities.
Note that free-free emission (thermal bremsstrahlung)
and radiative-Compton scattering
effectively erase any distortions\cite{Danese82}\cite{Sarkar84},\cite{Ellis92}
to a Planckian spectrum for epochs earlier than $z \sim 10^7$.

\begin{figure}
\centerline{\epsfxsize=9 cm\epsfbox[140 236 452 472]{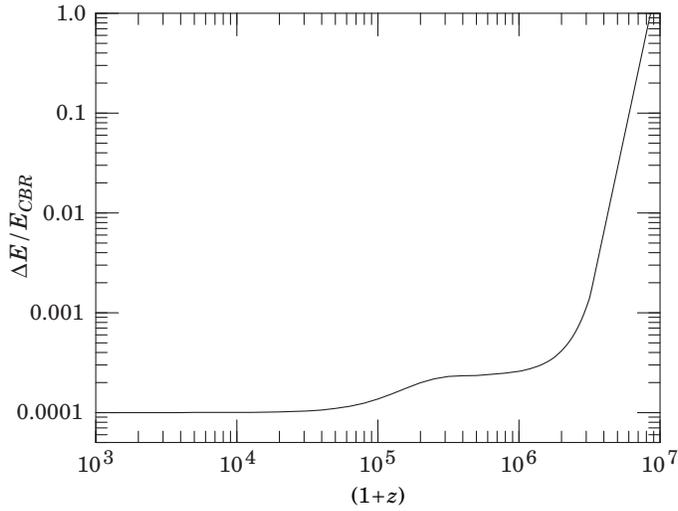}}
\caption{Upper Limits (95\%\ CL) on fractional energy ($\Delta E/E_{\rm CBR}$) releases
as set by lack of CMB spectral distortions resulting from
processes at different epochs.  These can be translated into constraints
on the mass, lifetime and photon branching ratio of unstable relic
particles, with some additional dependence on cosmological parameters
such as $\Omega_B$ [55][21]. }
\end{figure}

\subsection{Free-free distortion} {\it Very late energy release ($z\ll10^3$).}
Free-free emission from recent reionization $(z < 10^3)$ and
from a warm intergalactic medium 
can create rather than erase spectral distortion
in the late universe. 
The distortion arises
because of the lack of Comptonization at recent epochs.
The effect on the present-day CMB spectrum is described by
\begin{equation}
\Delta T_{f\!f} = T_{\gamma}\; {Y_{f\!f}}/{x^2},
\end{equation}
where $T_{\gamma}$ is the undistorted photon temperature,
$x$ is the dimensionless frequency, and
$Y_{f\!f}/x^2$ is the optical depth to free-free emission:

\begin{equation}
Y_{f\!f}
= \int^z_0 ~\frac{T_e(z') - T_{\gamma}(z') }{ T_e(z') }
\frac{ 8 \pi e^6 h^2 n_e^2 \;g }{ 3 m_e (kT_{\gamma})^3\;
 \sqrt{6\pi\, m_e\, k T_e} }
\;\frac{dt}{dz'} dz'\period
\end{equation}
\noindent
Here $h$ is Planck's constant,
$n_e$ is the electron density and $g$ is the Gaunt
factor\cite{bartlett91}.

\section{Spectrum Observations}
Beginning with the original discovery of the CMB by Penzias and Wilson
there was a rush of observations in the period 1965 through 1967.
For the most part interest shifted to the short-wavelength (high-frequency)
portion of the spectrum to observe the peak and Wien turn down 
to show that the spectrum was thermal. In the early 1980's effort
was renewed for observing the low frequency (Rayleigh-Jeans) region
primarily by a USA-Italian collaboration consisting of my group (LBNL-Berkeley),
Bruce Partridge (Haverford), Reno Mandolesi's group (Bologna), and 
Giorgio Sironi's group (Milano) with theoretical support from 
Luigi Danese and Gianfranco DeZotti (Padua).
We determined that scientific goals and technology had advanced 
to the point that the long wavelength ($>$ 1~cm) region should and
could be measured more accurately.

The general experimental concept is to observe the total power coming
from the sky and compare that to a well known reference source
using specially designed radio receivers called radiometers.
The references we used were carefully designed blackbodies with
total temperature of about 3.8~K which was very close to the total sky
signal so that the gain calibration of our radiometers was not 
a critical component of the observation.
The sky signal is the sum of many terms
\begin{equation}
T_{sky} = T_{CMB} + T_{atmosphere} + T_{Galaxy} + T_{Sun/Moon} + T_{instrum}
          + T_{terrestial}
\end{equation}
The total sky signal is dominated by the CMB, $T_{CMB} = 2.73$~K, and 
the atmospheric emission, $T_{atmosphere} \approx 1$~K. 
The other terms are generally
much smaller giving a total sky signal close to 3.8~K of total power 
which closely matches the absolute reference load.
(At very low frequencies the Galactic signal tends to be greater but
the atmospheric component decreases.)

The signal from the atmosphere is determined by scanning the radiometer
beam through various zenith angles and thus varying the air mass observed
and the data are then constrained by continuity and comparison to an
atmospheric model.

Our groups carried out a series of measurements from the high, dry sites
at White Mountain University of California Research Station (3800 meters)
which is in the rain shadow of the Sierras and the NSF South Pole Station.
Table 4 
at the end of the text (since it is so large) 
gives a summary of the low-frequency observations
and observations excluding those from {\it COBE} FIRAS and the UBC
rocket borne experiment which are in separate tables.

\subsection{Interstellar Molecules and Atomic Systems}
Observations of interstellar molecules and atomic systems,
most especially cyanogen (CN), provide a probe of the CMB temperature
in narrow wavelength bands at remote locations.
CN has proved most used and most precise as the observations are 
made at optical wavelengths and use well-developed technology.
They are by their nature indirect observations in that what is
measured is the relative population of various energy levels
and the ratio can be used to estimate $T_{CMB}$.
Thus one is observing the total excitation of the system
and accounting must be done to subtract the contribution of other
potential sources.
Fortunately, for cold, non-dense clouds, the contribution from sources
other than the CMB tends to be quite small, typically on the order of 0.1~K
compared to 2.73~K.

CN molecules existence abundantly in interstellar clouds.
If a cloud lies along the line of sight from a bright optical source,
the CN produces narrow absorption lines features on the source spectrum.
These absorbtion lines were detected in 1941 (more than half a century ago
and 23 years before the CMB was discovered by Penzias and Wilson)
by Adams \cite{Adams1941}, McKellar \cite{McKellar1941}, and others
and were noted as a puzzle. McKellar estimated a value for the excitation
temperature of the CN of a few degrees.
Immediately after the CMB discovery many people began observations of 
CN systems for information about the CMB temperature. Those efforts
have continued to the present and the most recent results are shown 
in the following table along with some results for other molecules.
Note that there is a discrepancy 
of results on CN between \cite{Palazzi92} and \cite{Roth95}.

\begin{table}[htb]
\begin{center}
\caption{Recent Molecular Measurements of $T_{CMB}$}
\begin{tabular}{lccr}
\hline 
Reference & Molecule & Wavelength & Temperature \\
 (Year)   &          &    (mm)    &   (K)        \\
\hline
Meyer \& Jura (1985) & CN & 2.64 & $2.70 \pm 0.04$~K \\
Meyer et al.  (1989) & CN & 2.64 & $2.75 \pm 0.03$~K \\
Meyer et al.  (1989) & CN & 2.64 & $2.77 \pm 0.07$~K \\
Meyer et al.  (1989) & CN & 2.64 & $2.75 \pm 0.08$~K \\
Meyer \& Jura (1985) & CN & 1.32 & $2.76 \pm 0.20$~K \\
Meyer et al.  (1989) & CN & 1.32 & $2.83 \pm 0.09$~K \\
Meyer et al.  (1989) & CN & 1.32 & $2.85 \pm 0.16$~K \\
Crane et al.  (1986) & CN & 2.64 & $2.74 \pm 0.05$~K \\
Crane et al.  (1989) & CN & 2.64 & $2.796^{+0.014}_{-0.039}$~K \\
Palazzi et al.(1990) & CN & 1.32 & $2.83 \pm 0.07$~K \\
Palazzi et al.(1990) & CN & 1.32 & $2.82 \pm 0.11$~K \\
Kaiser \& Wright (1990) & CN & 2.64 & $2.75 \pm 0.04$~K \\
Palazzi et al.(1992) & CN & 2.64 & $2.817 \pm 0.02$~K \\
Roth et al.   (1993) & CN & 2.64 & $2.729^{+0.023}_{-0.031}$~K \\
Roth \& Meyer (1995) & CN & 2.64 & " \\
\hline
Thaddeus (1972)      & CH & 0.76 & $< 5.23 $~K \\
Thaddeus (1972)      & CH$^+$ & 0.36 & $< 7.35 $~K \\
Kogut et al. (1990)  & H$_2$CO & 2.1 & $3.2 \pm 0.9$~K \\
\hline
Weighted mean $\pm 1 \sigma$ & & & $2.76 \pm 0.03 $~K \\
\hline
\end{tabular}
\end{center}
\end{table}

\subsection{{\it COBE} FIRAS}
It is the measurements of the FIRAS (Far-InfraRed Absolute Spectrophotometer)
on the {\it COBE} satellite that have shown definitively that the CMB
spectrum is very nearly Planckian.
The FIRAS instrument is a twin-input, twin-output polarizing Michelson
interferometer that achieves high precision by making a differential
rather than an absolute measurement.

One input is connected to view the sky through a large, low side-lobe sky horn.
The other input is connected to an internal 
calibrator at all times.
The internal calibrator is nearly a blackbody (96-98\%\ emissivity)
over the full wavelength range and is very stable.
The calibrator temperature is adjusted to give nearly null interferometer output.
\begin{figure}
\centerline{\epsfxsize=8 cm\epsfbox[-77 8 359 342]{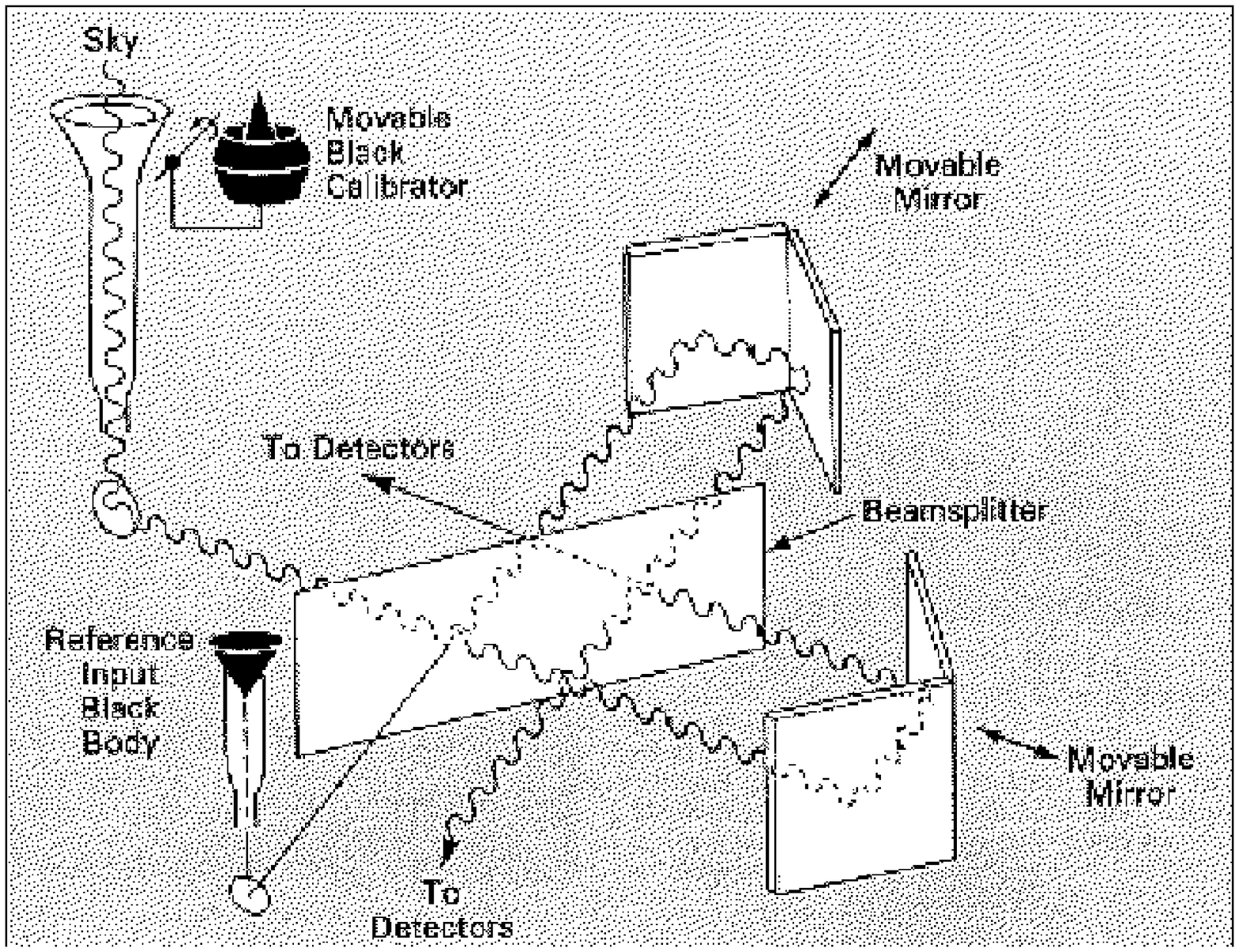}}
\caption{Schematic of the FIRAS instrument. There are two symmetric inputs:
the power from the sky and from an internal reference blackbody.
A high precision blackbody can be substituted for the sky signal input.
It is a polarizing Michelson interferometer with two dihedral mirrors.}
\end{figure}

The sky horn can be filled by the external calibrator
by swinging it on its pivot.
The external calibrator is a re-entrant absorbing cone. 
The combined external calibrator and sky horn cavity is a very good blackbody
with emissivity measured to be greater than 99.99\%\ and calculated
to be greater than 99.999\%.
The external calibrator temperature is commandable and 
was adjusted around null defined by the sky signal 
to provide an absolute and relative calibration.
This operation is possible since one does not have to be concerned
with windows or freezing of the atmosphere on the instrument 
and calibrator or with serious thermal loading.

Comparison of the signal from the sky with the signal from the external
calibrator with temperature adjusted to match gives an accurate
and precise measurement of deviations of the sky spectrum from a blackbody.
When these small deviations are added to the calculated Planck spectrum,
the FIRAS observed spectrum is produced.
See Figure 5 
for the measured spectrum and a 2.728~K Planckian.
\begin{figure}
\centerline{\epsfxsize=8 cm\epsfbox[120 270 500 620]{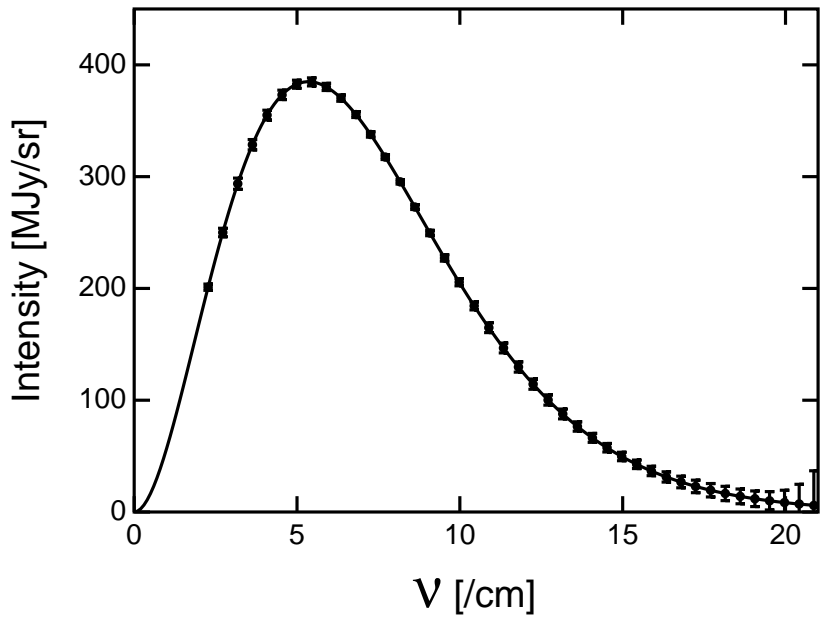}}
\caption{The spectrum of the CMB measured by FIRAS. The error bars shown
are $\pm 400 \sigma$. The solid line is a 2.728~K blackbody.}
\end{figure}
The temperature of the external calibrator, when the output
matches the sky viewing output, is the sky temperature.
A number of small corrections must be made, e.g. to the GRT 
(germanium resistance thermometers) readings, cosmic ray hits,
extra signal from interstellar dust or the experiment.
Another method is to use the wavelength of the peak of the brightness spectrum
determined by the length scale set by the dimensions of the interferometer 
and which is accurately checked and calibrated by the molecular lines 
observed in the Galactic emission by the interferometer.
A third approach is to use the dipole spectrum (see dipole spectrum section)
to set the temperature scale.

\begin{table}[htb]
\begin{center}
\caption{{\it COBE} Measurement of $T_{CMB}$}
\begin{tabular}{cr}
\hline 
Method & Temperature \\
\hline
GRT at sky match & $2.730 \pm 0.001$~K  \\
Peak of $\partial B_\nu / \partial T$ & $2.726 \pm 0.001$~K \\
FIRAS Dipole Spectrum & $2.717 \pm 0.007$~K \\
DMR Annual Dipole & $2.725 \pm 0.015$~K \\
\hline
Weighted mean $\pm 1 \sigma$ & $2.728 \pm 0.002$~K \\
\hline
\end{tabular}
\end{center}
\end{table}

Since the RMS deviation of the spectral intensity $I_\nu$
from a blackbody $B_\nu$ is $5 \times 10^{-5}$ of the peak $B_\nu$ amplitude,
the Planck function must be subtracted before plotting,
for residuals to be seen (e.g. see Figure 6). 
\begin{figure}
\centerline{\epsfxsize=9 cm\epsfbox[130 210 480 540]{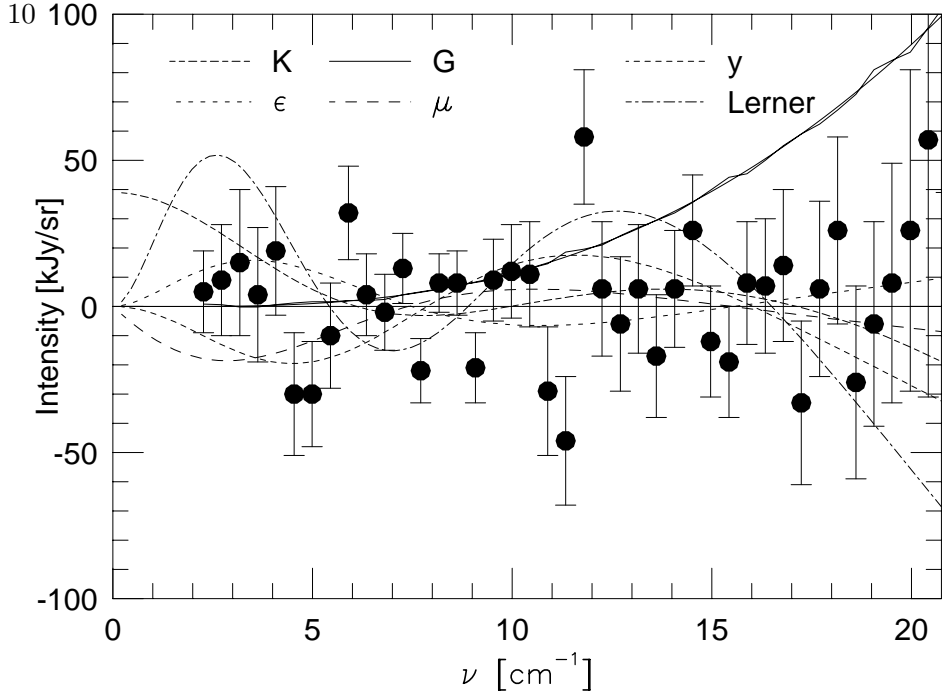}}
\caption{The residuals after subtracting a 2.728~K blackbody
and the best fitted galactic emission 
plotted at their 95\%\ confidence interval limit amplitudes.
The distortions are shown in the same manner as the data:
they are the residuals from a least squares fit to a temperature change
and the best-fitted galactic emission amplitude. 
The solid line (G) shows the observed galactic emission
and the smooth model fitting it. 
The curve K shows a constant intensity, $I_\nu$, distortion. 
The curve labeled $\epsilon$ shows an emissivity different from unity.
The curve labeled $\mu$ shows a Bose-Einstein (chemical potential) distortion.
The curve labeled $y$ shows a Compton distortion.
The curve labeled ``Lerner'' is a fit to the 1994 FIRAS data 
combining the $\epsilon$ and $y$ distortions which represented an effort
to have a `plasma' model that explained the data; but, it is a poor
fit to the improved data. }
\end{figure}
In fact the data are fitted to a form
\begin{equation}
I_\nu = B_\nu(T_0) + \Delta T \frac{\partial B_\nu}{\partial T} + g G(\nu)
\end{equation}
where $G(\nu)$ is the observed spectrum of the Galactic emission and the 
parameters $\Delta T$ and $g$ are adjustable to allow for a temperature
correction and an unknown amount of residual Galactic emission in the
darkest parts of the sky.

Using the FIRAS measured spectrum or deviations one can fit for distortions
and find the results in the following table:
\begin{table}[htb]
\begin{center}
\caption{FIRAS Spectral Distortion Limits}
\begin{tabular}{lcc}
\hline 
Distortion& Best Fit $\pm \sigma$ & 95\%\ CL Limit \\
\hline
$y$   & $(-1 \pm 6) \times 10^{-6}$ & $1.5 \times 10^{-5}$\\
$\mu$ & $(-1 \pm 4) \times 10^{-5}$ & $9 \times 10^{-5}$\\
$\epsilon -1$ & $(1 \pm 5) \times 10^{-5}$ & $11 \times 10^{-5}$\\
$\Delta I_\nu$ & $(9 \pm 15)$ kJy/sr & 39 kJy/sr\\
\hline
\end{tabular}
\end{center}
\end{table}
The first two distortions are the Compton and chemical potential distortions
discussed above. The next is allowing for an emissivity different than unity.
It is clear that the CMB is extremely close to the blackbody thermal spectrum.
The last $\Delta I_\nu$ allows for an offset either from the sky or
the instrument.

FIRAS also measures the spectrum of the dipole anisotropy which is shown here
but is discussed in the dipole spectrum section.
\begin{figure}
\centerline{\epsfxsize=8 cm\epsfbox[115 110 500 500]{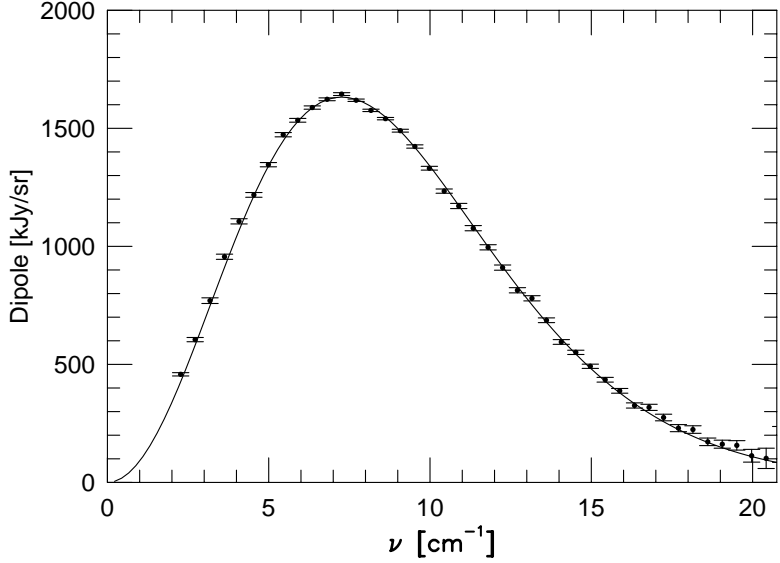}}
\caption{The spectrum of the CMB dipole as measured by FIRAS. 
The solid line is the derivative of a T = 2.728~K Planck function.}
\end{figure}
\begin{figure}
\centerline{\epsfxsize=9 cm\epsfbox[115 100 522 420]{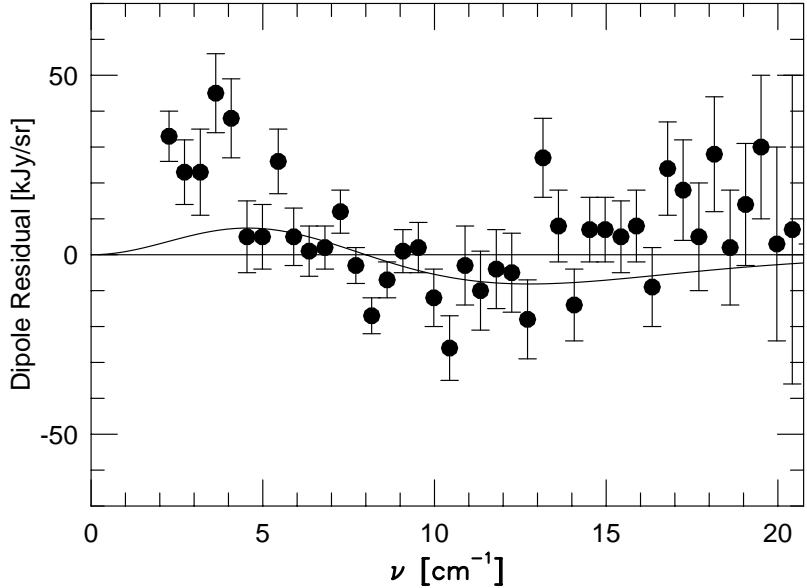}}
\caption{The CMB dipole spectrum residuals measured by FIRAS
after subtracting the derivative of the T = 2.728~K Planck function.
The curve shows the best fit letting both the dipole amplitude and 
CMB temperature vary.}
\end{figure}

\section{Spectrum summary}
The CMB spectrum is consistent with a blackbody spectrum over
more than three decades of frequency around the peak.
A least-squares fit to all CMB measurements yields:


\bigskip
%
 \begin{tabbing}
     p'=Effec pressure \=   \kill
\> $T_{\gamma} = 2.73 \pm 0.01$ K \\
\> $n_{\gamma} = (2\zeta(3)/\pi^2) T_\gamma^3 \simeq 413\,
 {\rm cm^{-3}}$ \\
\> $\rho_{\gamma} = (\pi^2 /15) T_\gamma^4 \simeq 4.68
 \times 10^{-34}\,
        {\rm g}\,{\rm cm^{-3}} = 0.262\,{\rm eV}\,{\rm cm^{-3}}$\\
\> $|y| < 1.5 \times 10^{-5}$ ~~~~~(95\%\ CL)\\
\> $|\mu_0| < 9 \times 10^{-5}$ ~~~~~~(95\%\ CL)\\
\> $|Y_{f\!f}| < 1.9 \times 10^{-5}$ ~~(95\%\ CL)\\
  \end{tabbing}
\bigskip

\noindent
The distortion parameter limits quoted here correspond to 
limits \cite{mather94}, \cite{bern94} on energetic processes
$\Delta E / E_{\rm CBR} < ~2 \times 10^{-4}$
occurring between redshifts $10^3$ and $5 \times 10^6$
(see Figure 3. 
The best-fit temperature from the COBE FIRAS
experiment is $T_\gamma=2.728\pm0.002\,$K\cite{fixsen96}.

\section{Spectrum Interpretation \& Discussion}

\subsection{Significance of CMB Being Planckian} 
Possibly the strongest arguments for the Big Bang model
are the CMB's existence and particularly its Planckian nature. 
This means that the CMB is both very cold and highly thermalized.
Since there are roughly $10^9$ photons to each baryon in the Universe, 
it is very difficult to produce the CMB in astrophysical processes
such as the absorption and re-emission of starlight by cold dust
(even iron needles) or the absorption or emission by plasmas.

All alternative models and modifications to the simplest big bang model
produce distortions to the CMB spectrum that have a $y$ component.
It is interesting to note that any deviation from a perfectly homogeneous,
isotropic, and isochronous universe causes a spectral distortion.
This is a result of the fact that the sum of two blackbody spectra 
of different temperatures does not result in a blackbody spectrum.
In the form discussed above a $y$ distortion is simply the convolution
of Planckian spectra. 

Thus for example, although the energy content of the CMB is comparable
to that in starlight and it is possible that dust absorption, processing, and 
re-emission could shift the radiation frequency to this range, it is extremely 
unlikely that the sum of all this radiation would just match a Planckian.
If somehow the dust were optically thick on cosmological scales, 
it is still not possible that the sum of red shifted emission from each
shell would add to a Planckian for all observers. 
Full arguments for dust or plasma filled universes must make use of
additional observations but in general there is an inconsistency with 
being able to see distant extragalactic sources at many wavelengths,
the observed CMB spectrum, and the Copernican Principle.

Likewise, this means that for all angular scales less than 
the FIRAS beam size of 7$^\circ$,
rms anisotropies cannot exceed about $\Delta T/ T < 10^{-3}$,
otherwise the superposition of temperatures would produce a 
$y > 10^{-5}$.

\subsection{Knowledge of $T_{CMB}$}
The CMB temperature, $T_{CMB}$, is now known to a precision of 1\%.
This makes it the best known cosmological parameter.
If we assume that the CMB spectrum is blackbody, 
we can calculate the number of photons in the CMB:
\begin{equation}
n_{\gamma} = (2\zeta(3)/\pi^2) T_\gamma^3 
= 413 (T_{CMB} / 2.728~{\rm K})^3  \,{\rm cm^{-3}}
\end{equation}
It is a small change to include simple distortions provided we know
their value.
We can also compute the present energy density in CMB photons
\begin{equation}
\rho_{\gamma} = (\pi^2 /15) T_\gamma^4 \simeq 4.68 \times 10^{-34}\,
{\rm g}\,{\rm cm^{-3}} = 0.262\,{\rm eV}\,{\rm cm^{-3}}
\end{equation}
Since the temperature scales as $T_0 = T_i (1 + z_i)$,
we can calculate the photon number density, $n_{\gamma i} = (1 + z_i)^3 n_\gamma$,
and energy density, $\rho_{\gamma i} = (1 + z_i)^4 \rho_\gamma$,
for any epoch $i$ with redshift $z_i$.

In the early universe the CBR (cosmic background radiation) 
which is the cosmologically redshifted present day CMB radiation
dominated over the matter energy density and thus was critical
to the development of the Universe.
In addition most cosmological models and calculations, such as Big Bang
Nucleosynthesis, are done in terms of the CBR temperature or density.
In particular matter density is usually expressed in terms
of the ratio either to the critical density or to the CBR density.
E.g. BBNS gives the number density of baryons, $n_b$, as
\begin{equation}
n_b / n_\gamma = 2.3 \times 10^{-8} (\Omega_b h^2) = 5 \times 10^{-10} h^2 .
\end{equation}

There is also the effect of the CBR/CMB on high energy cosmic rays 
which depends primarily on the energy density and less so on the spectrum.
But the CMB implies a strong cut off of high energy protons at roughly
$10^{21}$~eV due to the photoproduction of pions.
Likewise, the existence of the CMB causes a cut off for high energy photons
(and electrons/positrons) due to electron-positron pair production
(compton scattering).

\subsection{Limits on Processes in the Early Universe}
There are many possible sources of energy release or augmentation
from processes occurring in the early universe, including
decay of primeval turbulence, elementary particles, cosmic strings, or
black holes.  The growth of black holes, quasars, galaxies, clusters,
and superclusters might also convert energy from other forms. 

\subsubsection{Early Generation of Stars and Reionization}
Wright \etal (1994a) also give limits on hydrogen burning following the
decoupling.  These results depend on using geometrical arguments (a
$\csc|b|$ fit) to estimate the maximum amount of extragalactic energy
that could have a spectrum similar to that of our own Galactic dust.  We
found a limit that is a factor of about 3 smaller than the polar
brightness of the Milky Way.  A better understanding of the Galactic
dust would help produce a tighter limit on these extragalactic signals.

Consider first population III stars liberating energy that is converted
by dust into far infrared light (using an optical depth of 0.02 per
Hubble radius), and assume that $\Omega_b h^2=0.015$.  In that case less
than 0.6\% of the hydrogen could have been burned after $z=80$.  As a
second example, consider evolving infrared galaxies as observed by the
IRAS.   For reasonable assumptions, we found that less than 0.8\% of the
hydrogen could have been burned in evolving IR galaxies.

We also obtained limits on the heating and reionization of the
intergalactic medium.  It does not take very much energy to reionize the
intergalactic medium, relative to the CMBR energy, 
because there are so few baryons relative to CMBR photons.  
Even the strict FIRAS limits permit a single reionization event 
to occur as recently as $z=5$.  More detailed
calculations by Durrer (1993) show that the energy required to keep the
intergalactic medium ionized over long periods of time is much more
substantial and quite strict limits can be obtained.  If the current $y$
limits were about a factor of 5 more strict, then it would be possible
to test the ionization state of the IGM all the way back to the decoupling.

If the IGM were hot and dense enough to emit the diffuse X-ray
background light, it should distort the spectrum of the CMBR by inverse
Compton scattering.  This is a special case of the Comptonization
process, with small optical depth and possibly relativistic particles.
Calculations show that a smooth hot IGM could have produced less than
10$^{-4}$ of the X-ray background, and that the electrons that do
produce the X-ray background must also have a filling factor of less than
10$^{-4}$.


\subsubsection{Limits on Primordial Anisotropy}
Primordial perturbations will undergo energy dissipation via Silk damping.
Energy released is more effective at short wavelengths
where there are more oscillations. Limits on energy release
are also limits on the primordial perturbation power spectrum. 
Hu, Scott, and Silk (1994) find an upper limit on the power spectrum
index of about $n = 1.55$.  
It is interesting that these calculations give tighter limits than existing
direct measurements, even though the spectrum is only an upper limit. 
These results are dependent on assuming that a power law is the correct
form for the fluctuations over 7 orders of magnitude of scale.  There is
little possibility of observational evidence to confirm this assumption
over such a wide range, since small scale fluctuations have long since
been replaced by nonlinear phenomena.

\subsubsection{Limits on Shear, Vorticity, Turbulence}

\subsubsection{Limits on Gravitational Energy from LSS formation}
Together, free-free and Comptonized spectra
can be used to detect the onset of nuclear fusion 
by the first collapsed objects.
Ultraviolet radiation from the first collapsed objects is expected to
photoionize the intergalactic medium.
Since these objects form by non-linear collapse of rare high-density peaks
in the primordial density distribution,
the redshift at which they form is a sensitive probe of the 
statistical distribution of density peaks 
and the matter content of the universe.
Various models \cite{Tegmark1994}, \cite{Liddle1995} of structure formation 
predict significant ionization
at redshifts ranging from $10 < z < 150$,
depending on the matter content and power spectrum of density perturbations,
with a ``typical'' value $z_{\rm ion} \approx 50$.

\subsubsection{Limits on Particle Decay}
Exotic particle decay provides another source for non-zero chemical potential. 
Particle physics provides a number of dark matter candidates, 
including massive neutrinos, photinos, axions, 
or other weakly interacting massive particles (WIMPs).  
In most of these models, the current dark matter 
consists of the lightest stable member of a family of related particles, 
produced by pair creation in the early universe.  
Decay of the heavier, unstable members to a photon or charged particle branch 
will distort the CMB spectrum provided the particle lifetime is greater 
than a year. 
Rare decays of quasi-stable particles 
(e.g., a small branching ratio for massive neutrino decay
$\nu_{\rm heavy} \rightarrow \nu_{\rm light} + \gamma$)
provide a continuous energy input, also distorting the CMB spectrum.
The size and wavelength of the CMB distortion are dependent upon the 
decay mass difference, branching ratio, and lifetime. 
Stringent limits on the energy released by exotic particle decay
provides an important input to high-energy theories
including supersymmetry and neutrino physics\cite{Ellis92}.

\subsubsection{Limits on Antimatter-matter mixing}
In baryon symmetric cosmologies matter-antimatter annihilations gives rise
to excessive distortions of the CMB spectrum \cite{Jones1978}.

\subsubsection{Limits on Primordial Black Hole Evaporation}
Only a very small fraction, $f = M_{planck} / M $, of matter can have formed
black holes in the mass range $10^{11} \leq M \leq 10^{13}$~gm
otherwise their evaporation in the epoch preceding recombination
would have resulted in excessive distortions.
For smaller blackholes the limit is much weaker, since for $M < 10^{11}$~gm,
evaporation would have taken place during the epoch when the photon
spectrum would be completely thermalized.
The constraints follow from the condition that no more than all the entropy
in the universe can come from blackhole evaporation so that
$f < 10^9 M_{planck}/M$.

\subsubsection{Limits on Superconducting Cosmic Strings \& Explosive Formation}
If they are to play an important role in large-scale structure formation,
superconducting cosmic strings would be significant energy sources,
keeping the Universe ionized well past standard recombination.
As a result, the energy input distorts the spectrum of the CMB
but the Sunyaev-Zel'dovich effect. The Compton-$y$ parameter attains a
maximum value in the range of $(1 - 5) \times 10^{-3}$ \cite{Ostriker87}.
This is well above the observed value.

Explosive models of large-scale structure formation must create
distortions in the CMB spectrum from the energy released in the shock waves.
The limits on Compton-$y$ parameter rule out explosive models
for structure on scales $> 15$~Mpc \cite{Levin92}.

\subsubsection{Limits on the Variation of Fundamental Constants}
Noerdlinger \cite{Noerd1973} pointed out that the intensity of the
Rayleigh-Jeans portion of the CMB spectrum gives the present values
of $kT$, independently of the value of the Planck constant, $h$,
while the wavelength at which the spectrum peaks gives $kT$ in 
combination with $h$. That the two temperatures agree
within errors imply that the variation of $h$ must not have exceeded
a few per cent since recombination ($z \sim 1000$).
Likewise a wide variety of $G$-varying cosmologies predict that
the CMB spectrum will follow the standard Planckian formula
multiplied by an epoch-dependent factor, which, in turn, is related
to $G(t)$ \cite{Narlikar1980}. The agreement between the brightness
temperature in the Rayleigh-Jeans region and the temperature
determined by the peak location constrain the possible variation
in the gravitational constant $G$. 
Likewise one can obtain limits on the variation in the cosmological constant
(energy density of the vacuum) \cite{Pollock1980}.
The shape of the spectrum also constrains the number of large spatial dimensions 
(taking into account the possibility of fractal dimensions) to very nearly
three ($\pm 0.02$).

\section{Future Observations \& Results}
FIRAS has done such a splendid job of measuring the spectrum for the
bulk of the CMB energy and at long wavelengths Galactic emission
is such a serious foreground, it is at first difficult to imagine
the motivation necessary to gather the resources for significant improvement. 
However, there are scientific motivations for improved measurements
and there are experiments that one can envisage that may make
worthwhile improvements in the observations of the CMB spectrum.

\subsection{Interstellar/extragalactic Molecules \& Atoms}
The use of interstellar molecules, such as CN (cyanogen), 
offer a probe of the CMB at a remote location.
There are two distinct potential scientific gains from such observations.
The first is demonstrating that the CMB is universal,
a thing that observations of the Sunyaev-Zeldovich effect also establishes
a little more indirectly.
The second is that the CMB temperature scales as $(1 + z)$ with redshift.
A number of indications that this might be the case exist
but I would not consider them to yet be definitive
(i.e. strong enough to rule out a model like the big bang).
The best direct upper limit is a measurement \cite{Songaila94} of the background
temperature in high-redshift primordial clouds from an experiment
aimed at measuring the primordial deuterium abundance.
The claimed direct measurement \cite{Songaila94a} 
is based upon measuring the relative populations of hyperfine states in 
neutral carbon atoms observed in a gas cloud at a redshift $z = 1.776$,
which indicate a thermodynamic temperature of $7.4 \pm 0.8$~K,
which is consistent with the big-bang prediction
$T(z) = (1 + z) 2.73~{\rm K}$ which is 7.58 K.

Another recent measurement by Ge et al. \cite{Ge97} has measured C I
again in a gas cloud at a redshift $z = 1.79$ with a result
of $T(z=1.97) = 7.9 \pm 1.0$~K at 0.61 mm. Scaling by 1+z, one finds
the Big Bang predicted value is 8.1~K which is again consistent.
With accumulating observations and understanding of excitation mechanisms
these measurements provide a definite tightening of allow region
for alternative cosmologies.

\subsection{ARCADE}
We consider that our long wavelength ground-based observations 
have come near the fundamental limits set by the atmosphere and
the galactic foreground. Only a very great effort or a large
space-based mission is likely to generate more than a very modest improvement.
At the very longest wavelengths a much better understanding of the Galactic
emission is required in order to make more than just a modest improvement.

However, at intermediate wavelengths - those in the centimeter (1-6 cm),
it is possible to improve the spectrum measurements significantly 
by balloon-borne or satellite-based experiments.
It takes a large effort and very precise measurements, including
careful control of systematics and very good absolute calibration,
to actually improve the various limits or measurements of distortion
parameters such as $\mu$ and $Y_{ff}$. But it is possible.

ARCADE (Absolute Radiometer for Cosmology, Astrophysics, and Diffuse Emission) 
is a balloon-borne instrument designed to make measurements
of the intermediate wavelength spectrum.
A conceptual schematic drawing of the instrument is shown in Figure 9.
\begin{figure}
\centerline{\epsfxsize=9 cm\epsfbox[29 120 582 643]{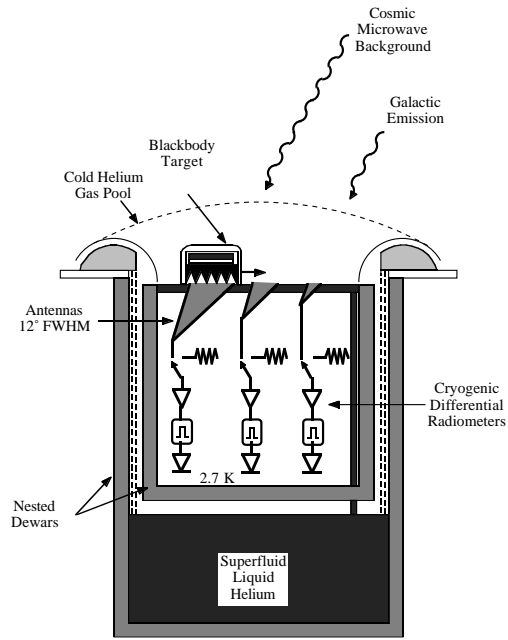}}
\caption{A schematic view of the ARCADE balloon-borne instrument}
\end{figure}
The instrument lives in a big bucket dewar, 
with a second dewar nested inside to allow
the aperture plane to remain cold even through it is nearly flush with
the mouth of the outer dewar.  
Fountain-effect pumps squirt superfluid liquid helium into a reservoir 
under the aperture plane assembly,
where it boils to keep the top plate cold (dumping the radiative heat load
from the IR lines in the atmosphere).  
Pinholes in the aperture plane vent the boiloff gas; 
a set of helium-cooled flares provide a bowl filled with a ``puddle" 
of cold helium gas.  
Provided the gas is colder than 20K, it's denser than ambient-temperature 
nitrogen and sits quietly as a transparent blocking layer between 
the cold optics and the warm atmosphere.  
The antennas are tipped 30 degrees with respect to the dewar symmetry axis, 
so that the dewar can remain upright (most of the time) 
while the antennas scan a circle 30 degrees in radius centered on the zenith.
The dewar tips occasionally to scan various atmospheric columns,
(i.e. different zenith angles to look through various amounts of atmosphere),
but this will be disruptive to the absolute target performance, 
so this happens only part of the time.
The anticipated measurement sensitivity is 1 mK from a balloon, 
limited by the ability to estimate/measure emission from the 
atmosphere, balloon, flight train, and Earth.  
ARCADE is basically a hardware 
development project for the eventual space mission.  
The design is kept such that the instrument can come off the balloon gondola 
and be put in a Spartan with minimal changes.

\subsection{DIMES}
The Diffuse Microwave Emission Survey (DIMES) 
has been selected for a mission concept study
for NASA's New Mission Concepts for Astrophysics program \cite{Kogut96b}.
DIMES will measure the frequency spectrum of the cosmic microwave
background and diffuse Galactic foregrounds at centimeter wavelengths to
0.1\% precision (0.1 mK), and will map the angular distribution to
20 $\mu$K per 6$^\circ$ ~field of view.  
It consists of a set of narrow-band cryogenic radiometers,
each of which compares the signal from the sky
to a full-aperture blackbody calibration target.
All frequency channels compare the sky to the {\it same}
blackbody target, with common offset and calibration,
so that deviations from a blackbody spectral {\it shape}
may be determined with maximum precision.
Measurements of the CMB spectrum complement CMB anisotropy experiments
and provide information on the early universe
unobtainable in any other way;
even a null detection will place important constraints
on the matter content, structure, and evolution of the universe.
Centimeter-wavelength measurements of the diffuse Galactic emission
fill in a crucial wavelength range
and test models of the heat sources, energy balance, and composition
of the interstellar medium.

\begin{figure}
\centerline{\epsfxsize=9 cm\epsfbox[-29 32 472 413]{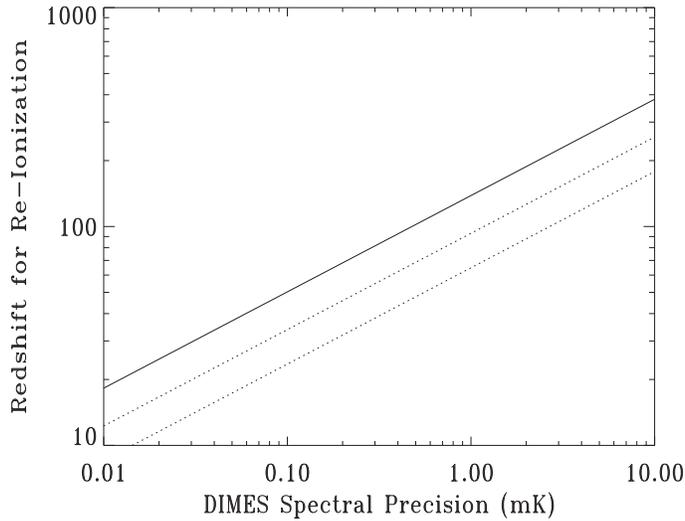}}
\caption[Redshift for reionized IGM]
{\small Upper limits to the redshift $z_{\rm ion}$ at which the
intergalactic medium (IGM) becomes reionized, 
as a function of the DIMES spectral precision.
The cosmologically interesting region $z_{\rm ion} < 100$
requires precision 0.1 mK or better.}
\label{dimes_reion} 
\end{figure}

The FIRAS measurement at sub-mm wavelengths shows no evidence
for Compton heating from a hot IGM.
Since the Compton parameter 
$y \propto n_e T_e$, 
the IGM at high redshift must not be very hot ($T_e \sim 10^5$ K)
or reionization must occur relatively recently ($z_{\rm ion} < 10$).
DIMES provides a definitive test of these alternatives.
Since the free-free distortion 
$Y_{\rm ff} \propto n_e^2 / \sqrt{T_e}$,  
lowering the electron temperature {\it increases} 
the spectral distortion \cite{bartlett91}. 
Figure 10 
shows the limit to $z_{\rm ion}$
that could be established from the combined DIMES and FIRAS spectra,
as a function of the DIMES sensitivity.
A spectral measurement at centimeter wavelengths with 0.1 mK precision 
can detect the free-free signature from the ionized IGM,
allowing direct detection of the onset of hydrogen burning.

\begin{figure}
\centerline{\epsfxsize=9 cm\epsfbox[-21 189 661 649]{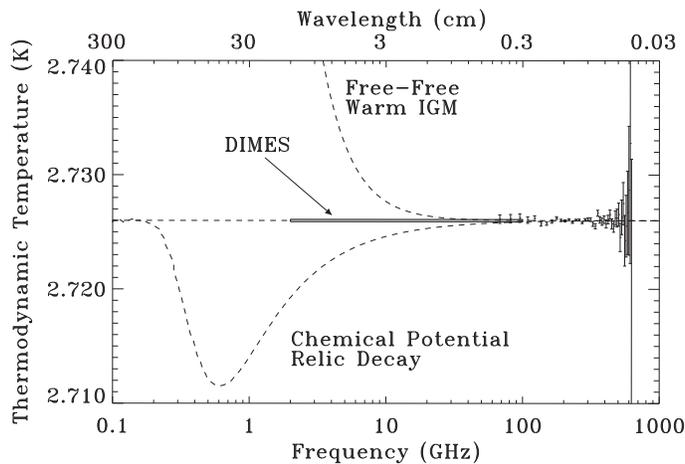}}
\caption[DIMES vs FIRAS Sensitivity]
{\small Current 95\% confidence upper limits to distorted CMB spectra.
The FIRAS data and DIMES 0.1 mK error box are also shown;
error bars from existing cm-wavelength measurements are larger than
the figure height.
An absence of distortions at millimeter and sub-mm wavelengths does {\it not}
imply correspondingly small distortions at centimeter wavelengths.}
\label{dimes_vs_firas}
\end{figure}

DIMES also provides a sensitive test for early energy releases,
such as the decay of exotic heavy particles or metric perturbations
from GUT and Planck-era physics.

DIMES will provide a substantial increase in sensitivity
for non-zero chemical potential (Figure 11). 
Such a distortion arises naturally in several models.
The {\it COBE} anisotropy data are well-described \cite{Gorski96} by a 
Gaussian primordial density field with power spectrum 
$P(k) \propto k^n$ per comoving wave number $k$, 
with power-law index $n = 1.2 \pm 0.3$.
Short-wavelength fluctuations which enter the horizon while the universe is
radiation-dominated oscillate as acoustic waves of constant amplitude and are
damped by photon diffusion, transferring energy from the acoustic waves to the
CMB spectrum and creating a non-zero chemical potential \cite{Daly92}, 
\cite{HuScottSilk94}.
The energy transferred, and hence the magnitude of the present
distortion to the CMB spectrum, depends on the amplitude of the perturbations
as they enter the horizon through the power-law index $n$.
Models with ``tilted'' spectra $n > 1$ 
produce observable distortions.

\subsubsection{Galactic Astrophysics}
Measurements of the diffuse sky intensity at centimeter wavelengths
also provide valuable information on astrophysical processes within our Galaxy.
Figure 12 
shows the relative intensity of 
cosmic and Galactic emission at high galactic latitudes.
Diffuse Galactic emission at centimeter wavelengths
is dominated by three components:
synchrotron radiation from cosmic-ray electrons,
electron-ion bremsstrahlung (free-free emission) 
from the warm ionized interstellar medium (WIM),
and thermal radiation from interstellar dust.
Despite surveys carried out over many years,
relatively little is known about 
the physical conditions responsible for these diffuse emissions.
Precise measurements of the diffuse sky intensity
over a large fraction of the sky,
calibrated to a common standard,
will provide answers to outstanding questions on physical conditions
in the interstellar medium (ISM):

\vspace{2mm}
$\bullet$~What is the heating mechanism in the ISM?  Is the diffuse gas 
heated by photoionization from the stellar disk, 
shocks, Galactic fountain flows,
or decaying halo dark matter?  

\vspace{2mm}
$\bullet$~How are cosmic rays accelerated?
Is the energy spectrum of local cosmic-ray electrons 
representative of the Galaxy as a whole?

\vspace{2mm}
$\bullet$~What is the shape, constitution, and size distribution of 
interstellar dust?  Is there a distinct ``cold'' component in the cirrus?

\begin{figure}
\centerline{\epsfxsize=9 cm\epsfbox[-6 199 638 650]{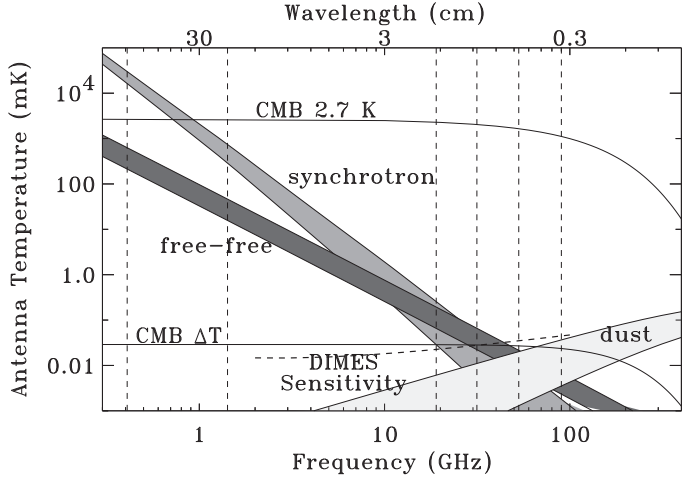}}
\caption[CMB and Galactic emission spectra]
{\small CMB and Galactic emission spectra.
The shaded regions indicate the signal range 
at high latitude ($|b| > 30^\circ$)
and include the effects of spatial structure 
and uncertainties in the spectral index of the Galactic emission components.
Solid lines indicate the mean CMB spectrum and rms amplitude of anisotropy.
Vertical dashed lines indicate existing sky surveys.
The DIMES sensitivity for a 6-month mission is shown.}
\label{foreground_spectra_fig}
\end{figure}

\vspace{2mm}
\noindent
The Galactic radio foregrounds
may be separated from the CMB by their frequency dependence
and spatial morphology.
DIMES will map radio free-free emission from 
the warm ionized interstellar medium.
The ratio of radio free-free emission to H$\alpha$ emission 
will map the temperature of the WIM to 20\% precision,
probing the heating mechanism in the diffuse ionized gas.
DIMES will have sufficient sensitivity to map the high-latitude
synchrotron emission,
probing the magnetic field and electron energy spectrum throughout the Galaxy.
Cross-correlation with the DIRBE far-infrared dust maps
will fix the spectral index of the high-latitude cirrus
to determine whether the dust has
enhanced microwave emissivity.

\subsubsection{Instrument Description}
Figure 13 
shows a schematic of the DIMES instrument.
It consists of a set of narrow-band cryogenic radiometers
($\Delta \nu / \nu \sim 10\%$)
with central frequencies chosen to cover
the gap between full-sky surveys at radio frequencies ($\nu < 2$ GHz)
and the {\it COBE} millimeter and sub-mm measurements.
Each radiometer measures the difference in power
between a beam-defining antenna (FWHM $\sim 6^\circ$)
and a temperature-controlled internal reference load.
An independently controlled blackbody target is located on the aperture plane,
so that each antenna alternately views the sky 
or a known blackbody.
The target temperature will be adjusted 
to null the sky-antenna signal difference
in the longest wavelength channel.
With temperature held constant,
the target will then move to cover the short-wavelength antennas:
DIMES will measure small spectral shifts about a precise blackbody,
greatly reducing dependence on instrument calibration and stability.
The target, antennas, and radiometer front-end amplifiers
are maintained near thermal equilibrium with the CMB,
greatly reducing thermal gradients and drifts.

\begin{figure}
\centerline{\epsfxsize=9 cm\epsfbox[ 30 150 582 761]{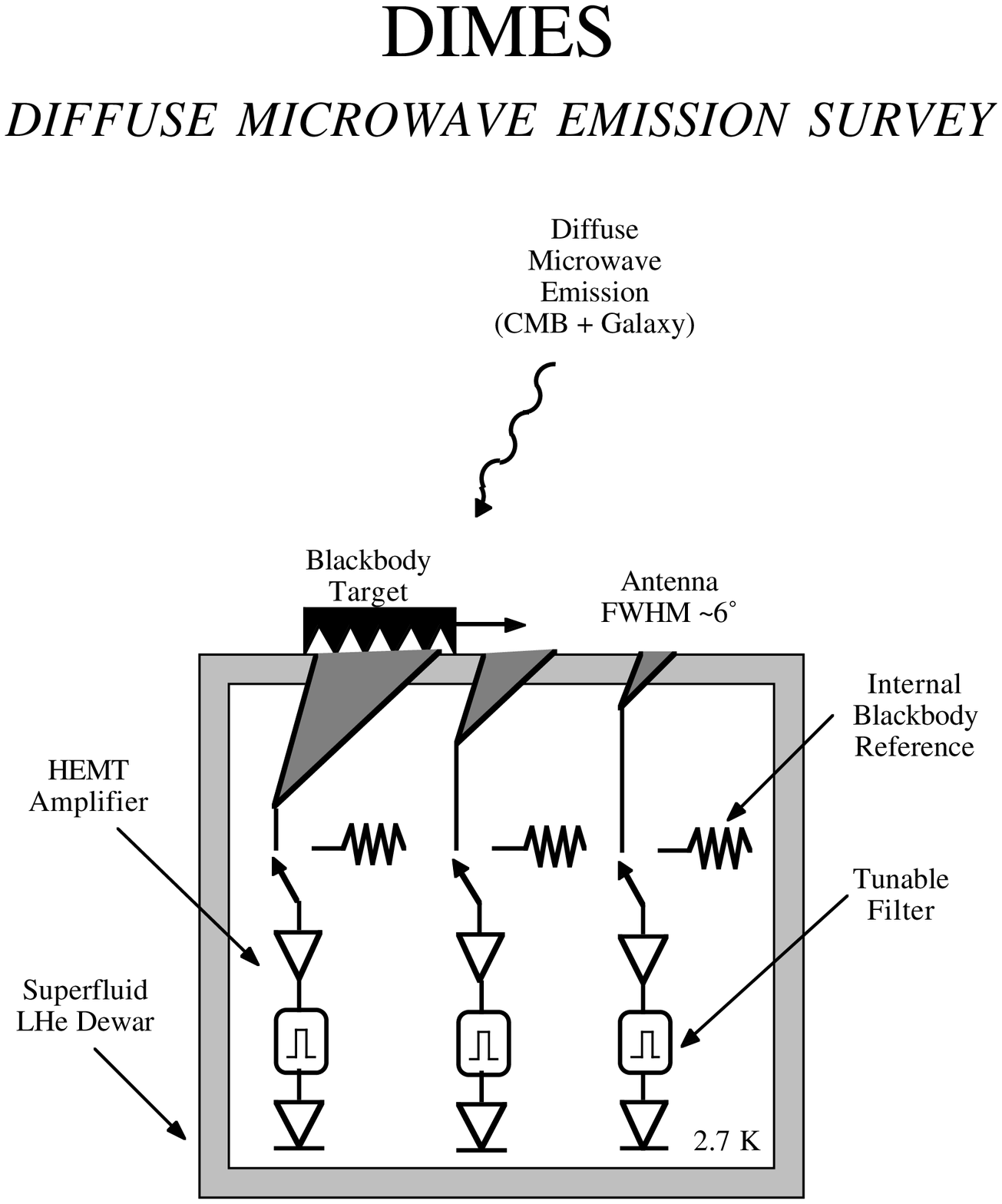}}
\caption[DIMES concept]
{\small Schematic drawing of DIMES instrument.}
\label{instrument_schematic}
\end{figure}

DIMES uses multiple levels of differences to reduce the effects of
offset, drifts, and instrumental signatures.
To reduce gain instability or drifts,
each receiver is rapidly switched 
between a cryogenic antenna
and a temperature-controlled internal reference load.
To eliminate the instrumental signature, 
each antenna alternately views the sky or a full-aperture target
with emissivity $\epsilon > 0.9999$.
To maximize sensitivity to spectral shape,
all frequency channels view the {\it same} target in progression,
so that deviations from a blackbody spectrum
may be determined much more precisely 
than the absolute blackbody temperature.

DIMES will remove the residual instrument signature
by comparing the sky to an external full-aperture blackbody target.
The precision achieved will likely be dominated by
the thermal stability of the target.
While the use of a single external target 
rejects common-mode uncertainties in the absolute target temperature, 
thermal gradients within the target or
variations of target temperature with time
will appear as artifacts in the derived spectra and sky maps.
Thermal gradients within the external target are reduced 
by using a passive multiply-buffered design
in which a blackbody absorber (Eccosorb CR-112, an iron-loaded epoxy)
is mounted on a series of thermally conductive plates with conductance $G_1$
separated by thermal insulators of conductance $G_2$.
Thermal control is achieved by heating the outermost buffer plate,
which is in weak thermal contact with a superfluid helium reservoir.
Radial thermal gradients at each stage 
are reduced by the ratio $G_2 / G_1$ between the buffer plates.
Typical materials (Fiberglass and copper) achieve 
a ratio $G_2 / G_1 < 10^{-3}$;
a two-stage design should achieve net thermal gradients 
well below 0.1 mK.
No heat is applied directly to the absorber,
and a conductive copper layer surrounds the absorber
on all sides except the front:
the Eccosorb lies at the end of an open thermal circuit,
eliminating thermal gradients from heat flow.

DIMES will not be limited by raw sensitivity.
HEMT amplifiers cooled to 2.7 K
easily achieve {\it rms} noise 1 mK Hz$^{-1/2}$,
reaching 0.1~mK sensitivity
in 100 seconds of integration.
The DIMES spectra are derived from comparison of the sky 
to the external blackbody target.
The largest systematic uncertainties arise from
thermal drifts or gradients within the target
and emission from warm objects outside the DIMES dewar (e.g., the Earth).
Thermometers buried in the microwave absorber 
monitor thermal gradients and drifts to precision 0.05 mK.
Emission from the Earth must be rejected at the -70 dB level to
avoid contributing more than 0.1 mK to the total sky signal.
DIMES will achieve this rejection using corrugated antennas with
6$^\circ$ ~beam and good sidelobe response;
two sets of shields between the aperture plane and the Earth
provide further attenuation of thermal radiation from the Earth.
{\it COBE} achieved -70 dB attenuation
with a 7$^\circ$ ~beam and a single shield \cite{Kogut96},
so the DIMES requirement should be attainable.

DIMES will eliminate atmospheric emission completely
by observing from low Earth orbit.
We are currently investigating the possibility
of utilizing the Spartan-400 carrier,
which will provide free-flyer capability 
to Shuttle orbits for 700 kg instruments
for a nominal mission of 6 to 9 months.

\section{Monopole Spectrum Summary}
The previous discussion reviews the observations, results, and future  
possiblities of the spectrum of the total CMB power.
In the next sections we consider the expected signal for a Planckian spectrum
for the monopole, dipole and higher order anisotropies
and how spectral distortions would appear in the frequency spectrum
of various anisotropies.

\section{Planckian Radiation Formula}
The specific intensity, $I_\nu$, of light
is defined as the incident energy per unit area, per unit solid angle,
per unit frequency.
\begin{equation}
I_\nu = \frac{2h\nu^{3}}{c^{2}} n
\end{equation}
where $h$ is Planck's constant, $\nu$ is the frequency, 
$c$ is the speed of light, and $n$ is the photon occupation number per mode.
The intensity or spectral brightness of a blackbody is a function of only one
parameter, the temperature
\begin{equation}
\label{eq:blackbody}
B_{\nu}(T)=\frac{2h\nu^{3}}{c^{2}}\frac{1}{e^{x}-1}
\end{equation}
where $x=h\nu/kT$. In the Rayleigh-Jeans region $x<<1$ and thus
\begin{equation}
\label{eq:RJblackbody}
B_{\nu}(T)=\frac{2\nu^{2}}{c^{2}}kT.
\end{equation}
The generalization of equation (\ref{eq:RJblackbody}) to any $x$
defines the antenna temperature of a blackbody
\begin{equation}
\label{eq:Tantdef}
B_{\nu}(T)=\frac{2\nu^{2}}{c^{2}}kT_{ant}(\nu).
\end{equation}
Rewriting equation (\ref{eq:Tantdef}) yields the relation between 
antenna and thermodynamic temperature 
\begin{equation}
\label{eq:Tant}
T_{ant}(\nu)= \frac{h\nu/k}{e^{x}-1}=T\frac{x}{e^{x}-1}.
\end{equation}
In the Rayleigh-Jeans portion of a blackbody spectrum the 
antenna temperature and the thermodynamic temperature are equal ($T_{ant}=T$).
%
%
Taking the derivative of equation (\ref{eq:Tant}) one obtains the relation 
between antenna and thermodynamic temperature {\it differences}
\begin{equation}
\label{eq:delTant}
\frac{ d T_{ant}}{d T} = \frac{x^{2}e^{x}}{(e^{x}-1)^{2}}
\end{equation}
where here $x=h\nu/kT_{o}$. 
The temperature difference conversion depends on
a knowledge of $T_{o}$ while equation (\ref{eq:Tantdef}) does not.
For example plugging 31.5, 53 and 90 GHz into equation (\ref{eq:delTant}) 
with $T_{o}=2.73$~K, 
we get the conversion factors 1.026, 1.074, 1.227 respectively.

\section{Dipole Formulae}
Observers with velocity $\vec{\beta} = \vec{v}/c$ through a Planckian
radiation field of temperature $T_{o}$ will measure directionally dependent
temperatures,
\begin{equation}
\label{eq:T}
T_{obs}(\theta)= T_{o}\frac{(1-\beta^{2})^{1/2}}{(1-\beta \mu)}
\end{equation}
where $\mu = cos\theta$ and $\theta$ is the angle between
$\vec{\beta}$ and the direction of observation as measured in the observer's
frame \cite{PeeblesWilkinson68}.We expand this through order $\beta^{3}$ to show that the dipole is the largest
member of a family of kinetic anisotropies,
\begin{equation}
\label{eq:Texpansion}
\frac{\Delta T}{T_{o}} =
\beta \mu +
\frac{\beta^{2}}{2}(2\mu^{2}-1) +
\frac{\beta^{3}}{4}(4\mu^{3}-2\mu)   + O(\beta^{4}).
\end{equation}
or
\begin{equation}
\label{eq:Texpansion}
\frac{\Delta T}{T_{o}} =
\beta cos \theta +
\frac{\beta^{2}}{2}cos2\theta +
\frac{\beta^{3}}{4}(4\mu^{3}-2\mu)   + O(\beta^{4}).
\end{equation}

The antenna temperatures of the CMB, the kinetic dipole and the normalizing
quadrupole amplitude are plotted in Figure 12. 

In the more general case of non-Planckian spectra $I_{\nu}$
we can define an equivalent antenna temperature by
\begin{equation}
\label{eq:general}
I_{\nu}= \frac{2\nu^{2}}{c^{2}}kT_{ant},
\end{equation}
which when combined with equation (\ref{eq:delTant}) yields
\begin{equation}
\frac{\Delta I}{I_{o}}=\frac{\Delta T_{ant}}{T_{ant}}
=\frac{\Delta T}{T_{o}}\frac{xe^{x}}{(e^{x}-1)},
\end{equation}
where $I_{o}$ is an isotropic but not necessarily Planckian radiation field
as seen by an observer in the rest frame of the field.

\section{The Dipole Anisotropy and Distortions of the CMB Spectrum}
The generalization of equation (\ref{eq:Texpansion}) for
motion through an isotropic but not necessarily Planckian radiation
field of  intensity $I_{o}(\nu)$ yields an observed intensity anisotropy,
\begin{equation}
\label{eq:I}
\frac{\Delta I}{I_{o}}(\nu ,\theta)=
\frac{I_{obs}(\nu ,\theta) - I_{o}(\nu )}{I_{o}(\nu )}.
\end{equation}
where $\nu$ is the frequency in the observer's frame
and $I_{o}$ is the intensity in the rest frame of the radiation.
The result to third order in $\beta$ is \cite{Lineweaver95}
\begin{eqnarray}
\nonumber
\frac{\Delta I}{I_{o}}&=&\beta \mu (3-\alpha_{1})
+ \frac{\beta^{2}}{2}\left[2\mu^{2}(6-3\alpha_{1} + \frac{\alpha_{2}}{2})
-(3-\alpha_{1}) \right]\\
&&+\frac{\beta^{3}}{4}
\left[4\mu^{3}(10-6\alpha_{1} + \frac{3}{2}\alpha_{2} - \frac{1}{6}\alpha_{3})
-2\mu(9-5\alpha_{1} + \alpha_{2})\right]+ O(\beta^{4})
\label{eq:Iexpansion}
\end{eqnarray}
where
$ \alpha_{n} = \frac{\nu^n}{I(\nu)}
 \frac{\partial^{n} I(\nu)}{\partial \nu^{n}}$.
A pedagogical check of this formula can be made by noticing that
for a Planckian spectrum
$\Delta I/I_{o}=\Delta T_{ant}/T_{ant}=
\frac{xe^{x}}{(e^{x}-1)}\Delta T/T_{o}$,
where $T_{ant}$ is antenna temperature and $x=h\nu/kT_{o}$.
In the Rayleigh-Jeans limit, $\alpha_{1}=\alpha_{2}=2$, $\alpha_{3}=0$
and one obtains $\Delta I/I_{o} =\Delta T/T_{o}$.
An analogous simplification does not occur in the Wien limit
because of the $\nu$ dependence of the derivatives $\alpha_{n}$.
Another check is that an $I \propto \nu^{3}$ non-Planckian spectrum
yields no kinetic anisotropy since I/$\nu^{3}$ is a Lorentz invariant.
For this case, $\alpha_{1}=3$ and $\alpha_{2}=\alpha_{3}=6$.

{\bf Summary}
The frequency dependence of the dipole anisotropy provides a means to 
determine the CMB temperature and to detect CMB spectral distortions.
In particular accurate measurements of the CMB dipole anisotropy
at multiple wavelengths may help in limiting or detecting small 
spectral distortions. On the other hand accurate spectral measurements
are needed for a precise interpretations of the observed anisotropy.
It is important to make measurements at as many wavelengths as possible.

\subsection{Introduction to Dipole Anisotropy Spectrum}
The dipole anisotropy has been measured well at many wavelengths, particularly
by the {\it COBE} DMR and FIRAS instruments. Prior to that several
experiments also measured the dipole anisotropy amplitude and direction.

The most obvious interpretation of the dipole anisotropy is in terms of the
peculiar velocity of the solar system; on the other hand it might result
from a combination of very long wavelength primordial perturbations
\cite{wilson&silk1980} \cite{Turner} \cite{Paczynski&Piran1990}.
We can certainly expect that on the order of 1\%\ of the dipole anisotropy
is due to primordial anisotropies based upon a simple extrapolation
of the observed anisotropy power spectrum.

Assuming that the observed dipole anisotropy results primarily from the
doppler shift due to the peculiar motion of the Solar System,
small spectral distortions must give rise via the Compton-Getting effect
to a characteristic frequency dependence of the dipole amplitude
arising from the shape of the spectral distortions.

\subsection{The Compton-Getting Effect}
The Compton-Getting effect is, in its original formulation
\cite{Compton&Getting1935}, the 24-hour variation in the cosmic ray intensity
due to the peculiar velocity of the Earth.
This effect is easily generalized as it a straight consequence of the Lorentz
invariance of the distribution functions of the particles and photons 
in phase space (see \cite{Forman1970} for a comprehensive discussion).

An observer with velocity {\bf v} ($\beta$ = v/c) with respect to 
the reference frame in which the photon distribution function $n(\nu)$
is isotropic to at least first order in $\beta$ 
will measure a difference between the intensity 
received in the direction of motion and that received in a direction
perpendicular to its motion proportional to:
\begin{equation}
\frac{\Delta n}{n} = \frac{d~ln n}{d~ln \nu} \beta
\end{equation}
Thus measurements of the dipole anisotropy of the CMB intensity
yield information on the slope of the spectrum.

To first order in $\beta$ 
the dipole anisotropy of the CMB intensity is
\begin{eqnarray}
T_d &=& \frac{h \nu}{k} [ 1/ln( 1 + 1/n(\nu) ) 
                     - 1/ln( 1 + 1/n(\nu[1+\beta]) ) ] \cr
&\approx& - \frac{h \nu}{k (1+n)} ln^{-2}( 1 + \frac{1}{n} )
 \frac{d~ln n}{d~ln \nu} \beta
\end{eqnarray}

In the case of a Planckian spectrum $[n = (exp(x) -1)^{-1}; x = h \nu/kT]$
the temperature anisotropy is independent of frequency and
\begin{equation}
\frac{T_d}{T}  \approx \frac{v}{c} = \beta .
\end{equation}

Deviations from a Planckian spectrum, however, lead to a dependence of the
dipole anisotropy amplitude, $T_d$, specific to the shape of the distortion.
Define $\delta$, the first order fractional change 
in the dipole anisotropy amplitude, to be 
\begin{equation}
\delta \equiv \frac{\Delta T_d}{T_{do}}  \approx [ \frac{T_d}{T} - \beta ] \beta^{-1} .
\end{equation}
Now we can calculate and plot the fractional change in dipole amplitude $\delta$
from predicted potential distortions.

\subsection{Application to Potential Distortions}
The three types of spectral distortions normally discussed are:
Comptonization distortion, Bose-Einstein distortion, and free-free distortion.
In addition it is sometimes pointed out that there are some very low level
distortions expected from the final stages of recombination.
Finally, it is possible that there is a generic distortion caused by
effects which have not been anticipated, calculated, or otherwise 
expected. We can make estimates of these also.

\subsubsection{Comptonization Distortion}
The first order approximation to the photon occupation number 
for a comptonized spectrum is
\begin{equation}
n_c = n_p [ 1 + u x exp(x) n_p (\frac{x}{tanh(x/2)} - 4)]
\end{equation}
where the parameter $u = k(T_e -T_\gamma)/m_e c^2$ 
is a measure of the amount of extra energy 
injected into the radiation field. 
Figure 14 
shows the dipole deviation, $\delta$ spectra
predicted for such distortions.
\begin{figure}
\centerline{\epsfxsize=9 cm\epsfbox[61 391 525 680]{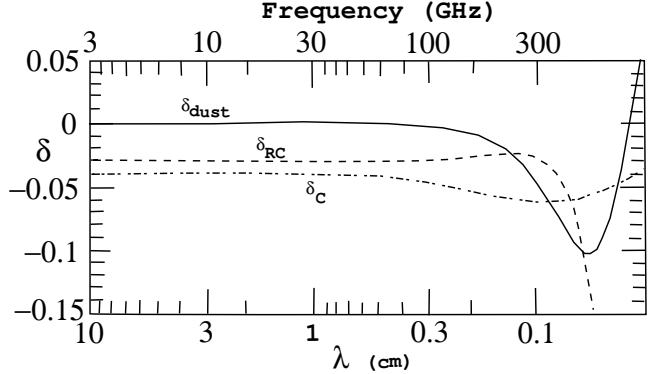}}
\caption{Predicted fractional variation of the dipole amplitude
( $\delta = [(T_d/T) - \beta]/\beta$ )
for models with non-relativisitc comptonization (C - dot-dashed line),
relativisitc comptonization (RC - dashed line),
and cold dust (dust - solid line) producing a sub-mm excess.
}
\end{figure}

\subsubsection{Bose-Einstein Distortion}
In a Bose-Einstein or chemical potential distortion
the photon occupation number~$n$ is 
\begin{equation}
n = \frac{1}{ e^{x + \mu_0} - 1 }\comma
\end{equation}
where  $x \equiv { h \nu / k T}$  and
$\mu_0$ is the dimensionless chemical potential. 
The chemical potential is predicted to be frequency-dependent,
\begin{equation}
\mu(x) = \mu_0 e^{- 2x_b/x }\comma
\end{equation}
where $x_b$ is the transition frequency at which
Compton scattering of photons to higher frequencies
is balanced by free-free creation of new photons.
The resulting spectrum has a sharp drop in brightness temperature
at centimeter wavelengths\cite{burigana91}
with a minimum at $\lambda_{\rm min}\simeq 4.5~\Omega_B^{-5/8}~{\rm cm}$.
Thus the minimum wavelength is determined by $\Omega_B$.

We can use this expression for the photon occupation number in 
the formula for the dipole anisotropy amplitude and find the 
fractional variation in the dipole anisotropy, $\delta$,
for the various possible values of energy release 
$\mu_0 \simeq 1.4 {\Delta E} / E_{\rm CBR}$
and other cosmological parameters, i.e. $\Omega_B$.

To first order the deviation is proportional to $\mu$.
\begin{equation}
\frac{T_d}{T} \approx \beta \frac{x^2}{(x + \mu)^2}
 (1 + \mu  \frac{2x_b}{x^2})
\end{equation}
where the second term in the parenthesis is generally small so that
so that
\begin{equation}
\delta = -\frac{2 \mu x + \mu^2}{(x + \mu)^2} 
= -2 \mu \frac{x + \mu/2}{(x + \mu)^2}
\end{equation}
\begin{figure}
\centerline{\epsfxsize= 9 cm \epsfbox[-88 88 705 701]{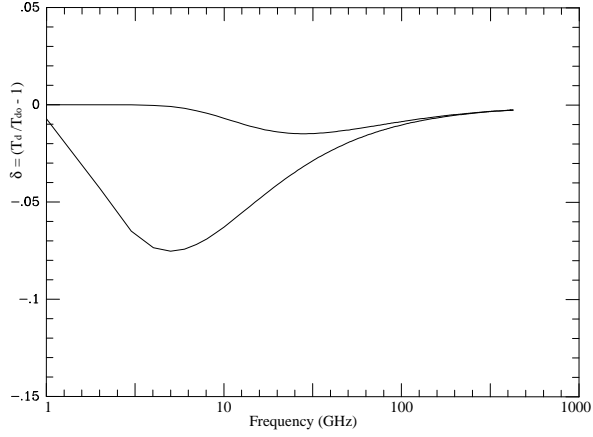}}
\caption{Fractional deviation in the dipole anisotropy amplitude
due to the doppler shift for two cases of a Bose-Einstein (chemical potential)
distortion
}
\end{figure}

\subsubsection{Free-Free Distortion}
Thermal bremsstrahulung from an ionized intergalactic medium
distorts the observed CMB spectrum changing the temperature by an amount
\begin{equation}
\Delta T_{f\!f} = T_{\gamma}\; {Y_{f\!f}}/{x^2},
\end{equation}
where $T_{\gamma}$ is the undistorted photon temperature,
$x$ is the dimensionless frequency, and
$Y_{f\!f}/x^2$ is the optical depth to free-free emission.
The predicted distortion is shown in Figure 16. 

\begin{figure}
\centerline{\epsfxsize= 9 cm \epsfbox[000 253 559 541]{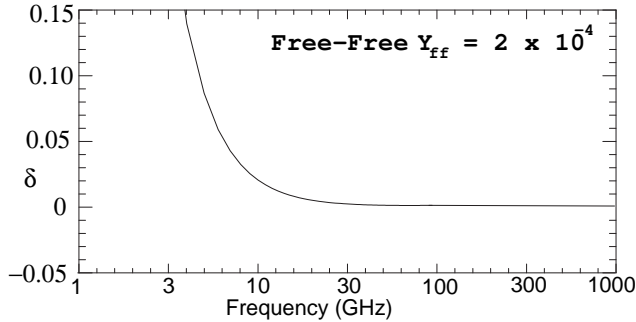}}
\caption{Fractional deviation in the dipole anisotropy amplitude
from a free-free emission distortion with $Y_{ff} = 2 \times 10^{-4}$.
}
\end{figure}

\subsubsection{Recombination Line Distortion}
Since there are on the order of $10^9$ CMB photons per baryon,
recombination does not have a large effect on the CMB spectrum.
However, there are small features ($10^{-4} - 10^{-8}$) that result
from atomic lines. 
The best known and calculated are the Lyman-$\alpha$ for hydrogen which
appear deep in the Wien region \cite{Peebles93}.
The result is a step at the high frequency side of the L$\alpha$ resonance
(divided by the redshift $z \lsim 1100$ of recombination). 
There is a slight smearing due to the natural line width set by atomic
parameters and the thermal motion.
The dominant effect is the cosmological expansion redshift 
which pulls photons from the low frequency side and deposits them
on the high frequency side. The final result is a slight step down
at the highest frequency at which the resonance was effective.

There are hydrogen resonances at lower frequencies, not only bound-free
but also bound-bound transitions, that are manifest in the radio and
mm wavelength range.
It appears difficult to detect these with an absolute measurement
even using a frequency switching system without a very substantial effort.
It is likely that using a narrow-bandwidth or spectral receiver
observing the dipole anisotropy is a more effective way to observe
such a line. 
The calibration of either such system requires a great deal of care.

\subsubsection{Unanticipated Distortions}
It is always possible that there are spectral distortions that do
not fall in the categories discussed above.
In particular, it is quite possible that astrophysical or particle
decay/interaction effects could alter the photon occupation number
at long wavelengths not yet measured precisely.

Although the precise {\it COBE} measurements carry implications
for possible distortions at longer wavelengths,
the absence of distortions near the peak CMB intensity does {\it not}
imply correspondingly small distortions at longer wavelengths.
Distortions as large as 5\% could exist 
at wavelengths of several centimeters or longer
without violating existing observations.

\section{SZ Measurements as a Probe of Spectral Distortions}
The Sunyaev-Zeldovich \cite{sunzel72} effect in the direction
of rich clusters of galaxies provides another probe of the CMB spectral
shape by means of differential measurements
(\cite{GouldRephaeli1978}; \cite{Fabbri1979}; \cite{Rephaeli1980};
\cite{Wright1983}; \cite{Rephaeli199x}).
The change in the CMB brightness temperature or intensity is 
essentially a second order Doppler effect.
The amplitude of the effect is proportional
to the second derivative of the intensity at the frequency of observation:
\begin{equation}
\frac{\partial I}{\partial y_C} = \frac{\partial^2 I}{\partial(ln x)^2}
- 3 \frac{\partial I}{\partial \ln x} \comma
\end{equation}
where $y_C$ is the comptonization parameter of the cluster
and $x = h\nu/kT_0$.
If the intensity ($I \propto x^\alpha$) is locally a power law with
exponent $\alpha$, then $\partial I / \partial y_C \propto \alpha (\alpha -3)$.
In the Rayleigh-Jeans region, $\alpha = 2$; it then decreases with increasing
frequency and becomes negative in the Wein region.
The Sunyaev-Zeldovich effect changes signs around the CMB spectrum peak.
The spectrum of the SZ effect is sensitive to the detailed shape
of the original CMB spectrum. 
Figure 17 
shows examples of the predicted effect.
\begin{figure}
\centerline{\epsfxsize= 9 cm \epsfbox[0 0 462 224]{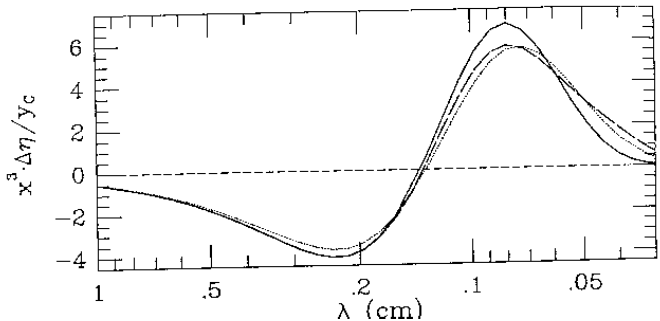}}
\caption{The predicted Sunyaev-Zeldovich effect corresponding to a 
Planck spectrum (solid line) and two spectral distortions:
non-relativisitc comptonization (dotted line) and a cold dust emission
(long dashes) creating a sub-mm excess.
For the ordinate $x = h\nu/kT; x^3 \Delta n$ is a quantity
proportional to the change of the CMB intensity, 
$y_C$ is the comptonization parameter for the cluster.}
\end{figure}
Two things, in addition to observational noise and errors, 
act to confuse. 
The first confusion is that the shape of the SZ effect is slightly
modified by the temperature distribution of the hot electrons
in the galactic cluster (\cite{Rephaeli199x}).
The second confusion is any local cluster or foreground emission contributions 
to the observed intensity. 
Fortunately, foreground emissions will not have a dipole pattern
or SZ effect and measurements of this kind can be used to separate
out extragalactic contributions to the observed flux.

\section{CMB Anisotropy Frequency Spectrum}
Given the precise observations of the monopole and dipole frequency spectrum,
then we can confidently predict the frequency spectrum of 
higher order CMB anisotropies.
The frequency spectrum should be the same as that for the
dipole anisotropy (except for the special case of the thermal SZ effect).
This is a fundamental assumption underlying techniques
for separating the observations of the microwave sky into its CMB
and foreground components.

We can ask, based on the COBE data, how well is this assumption verified.
It turns out tha FIRAS alone does not have sufficient resolution
to measure the higher order anisotropy frequency spectrum.
That is FIRAS can readily measure the dipole frequency spectrum
but is not able to measure that of the quadrupole, octopole, etc.
on its own.
However, if the FIRAS observations are crosscorrelated with the DMR
observations, then one can make an estimate of the anisotropy frequency 
spectrum \cite{fixsen97}.
This technique can and has been used with external experiments
such as FIRS, Tenerife, Saskatoon and will be with future observations;
but these other observations are currently much more limited
than FIRAS in frequency samplingng.

\begin{figure}
\centerline{\epsfxsize=9 cm\epsfbox[70 320 480 792]{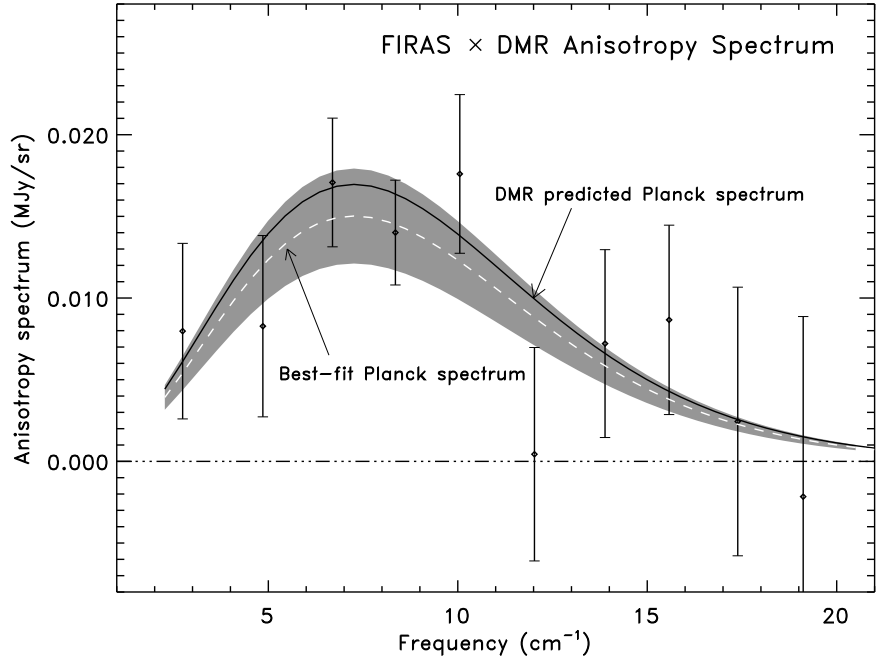}}
\caption{The frequency spectrum of primordial CMB distortions
determined by crosscorrelation of the COBE DMR and FIRAS data sets.
The points with errors are the results of correlating the DMR CMB 
anisotropy map with the FIRAS map generated by removing the
monopole, dipole, and an estimate of the Galactic dust emission.
The thin line is the predicted spectrum based upon the DMR data alone
and the assumption of a precisely Planckian CMB spectrum.
The band with centered line is the best-fitted Planckian CMB spectrum
to the FIRAS-DMR crosscorrelation.
}
\end{figure}

The observations of the CMB thermal spectrum and the frequency spectra
of anisotropies are point to a precisely thermal spectrum for the CMB.
This figure \cite{fixsen97} shows the three levels of frequency spectra: monopole,
dipole, higher order anisotropies left after Galactic dust emission is removed.
\begin{figure}
\centerline{\epsfxsize= 13 cm\epsfbox[0 75 612 792]{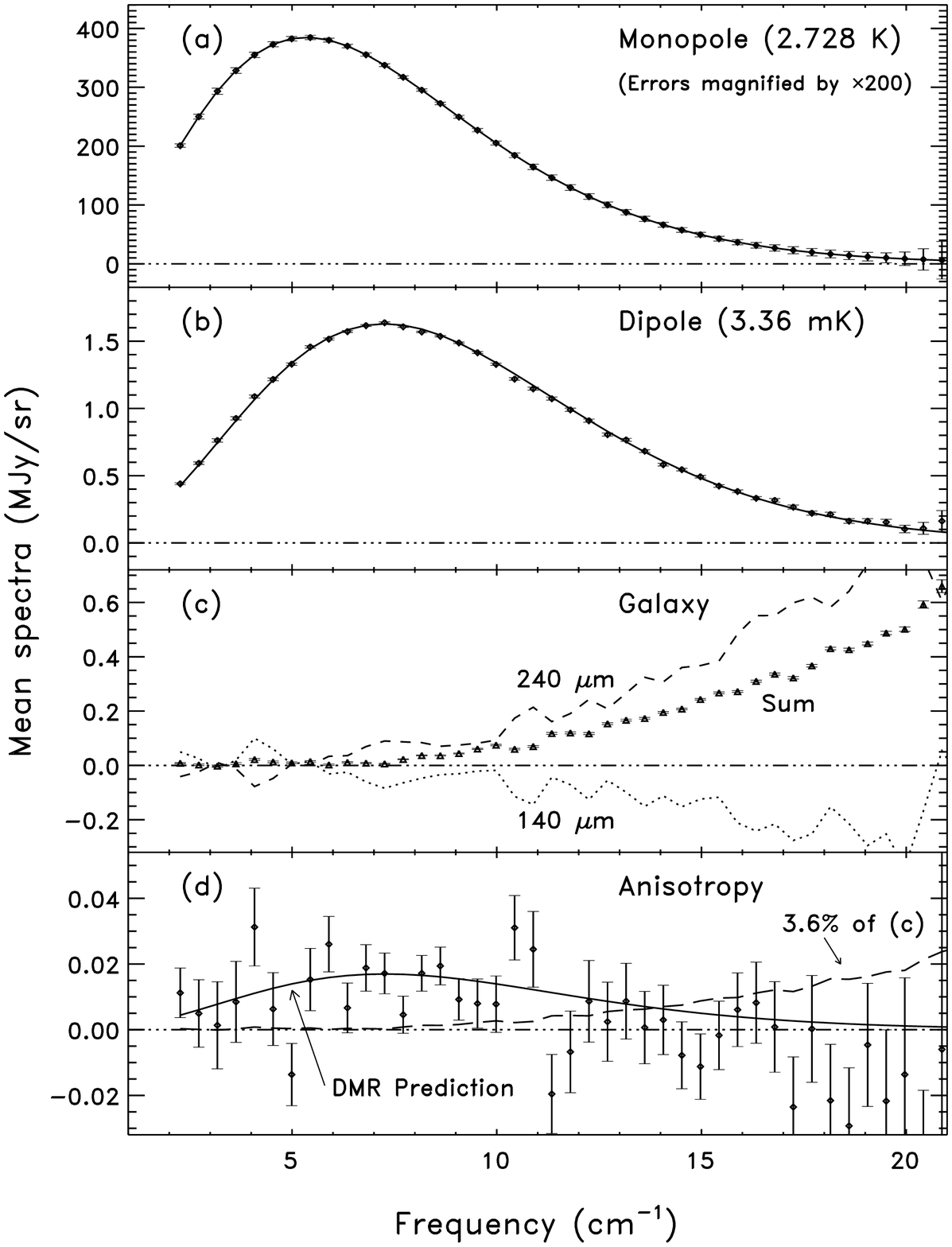}}
\caption{Frequency Spectrum of CMB features:
(a) The CMB monopole frequency spectrum as measured by FIRAS
and a line for a Planckian.
(b) The CMB dipole frequency spectrum as measured by FIRAS and a line 
indicating the anisotropy frequency spectrum expected for a Planckian.
(c) The frequency spectrum of Galactic dust emission removed 
from the FIRAS spectrum.
(d) The CMB higher order anisotropy frequency spectrum as measured by FIRAS and a line
indicating the anisotropy frequency spectrum expected for a Planckian.
}
\end{figure}

\section{Acknowledgments}
This work was supported in part
by the Director, Office of Energy Research, Office of High Energy and
Nuclear Physics, Division of High Energy Physics of he U.S. Department
of Energy under contract No. DE-AC03-76SF00098.

{}

\begin{table}[htb]
\begin{center}
\caption{Measurements of $T_{CMB}$}
\begin{tabular}{cccll}
\hline
Frequency & Wavelength & Temperature & Location & Reference\\
 (GHz)    &   (cm)     &   (K)       & (calibration) &     \\
\hline
0.408 &  73.5 & $3.7 \pm 1.2$ & Ground (LN)  & Howell \& Shakeshaft 1967, Nature, 216, 753.   \\
0.6   &  50   & $3.0 \pm 1.2$ & Ground (Term)& Sironi et al. 1990, Ap.J., 357, 301.  \\
0.610 &  49.1 & $3.7 \pm 1.2$ & Ground (LN)  & Howell \& Shakeshaft 1967, Nature, 216, 7  \\
0.635 &  47.2 & $3.0 \pm 0.5$ & Ground (LN)  & Stankevich et al 1970, Australian J. Phys, 23, 529  \\
0.820 &  36.6 & $2.7 \pm 1.6$ & Ground (Term)& Sironi et al. 1991, Ap.J., 378, 550.  \\
1     &  30   & $2.5 \pm 0.3$ & Ground (LN)  & Pelyushenko \& Stankevich 1969, Sov. Astron., 13, 223.  \\
1.4   &  21.3 & $2.11\pm 0.38$ & Ground (CLC)& Levin et al. 1988, Ap.J., 334,14 \\
1.42  &  21.2 & $3.2 \pm 1.0$ & Ground (Term)& Penzias and Wilson 1967, AJ, 72, 315  \\
1.43  &  21 & $ 2.65^{+0.33}_{-0.30}$ & Ground (LN) & Staggs et al. 1996, ApJ, 458, 407\\
1.44  &  20.9 & $2.5 \pm 0.3$ & Ground (LN)  & Pelyushenko \& Stankevich 1969, Sov. Astron., 13, 223. \\
1.45  &  20.7 & $2.8 \pm 0.6$ & Ground (Term)& Howell \& Shakeshaft 1966, Nature, 210, 1318. \\
1.47  &  20.4 & $2.27 \pm 0.19$ & Ground (CLC)& Bensadoun et al. 1992 (in press) \\
2     &  15   & $2.5 \pm 0.3$ & Ground (LN)  & Pelyushenko \& Stankevich 1969, Sov. Astron., 13, 223. \\
2.3   &  13.1 & $2.66 \pm 0.77$ & Ground (Term)& Otoshi \& Stelzreid 1975, IEEE Trans on Inst \& Meas, 24, 174. \\
2.5   &  12   & $2.71 \pm 0.21$ & Ground (CLC) & Sironi et al. 1991, Ap. J., 378, 550. \\
3.8   &  7.9  & $2.64 \pm 0.06$ & Ground (CLC) & De Amici et al. 1991, Ap.J., 381, 341. \\
4.08  &  7.35 & $3.5  \pm 1.0 $ & Ground (Term)& Penzias \& Wilson 1965, Ap.J., 142, 419. \\
4.75  &  6.3  & $2.70 \pm 0.07$ & Ground (CLC) & Mandolesi et al. 1986, Ap.J., 310, 561. \\
7.5   &  4.0  & $2.60 \pm 0.07$ & Ground (CLC) & Kogut et al. 1988, Ap.J., 355, 102\\
7.5   &  4.0  & $2.64 \pm 0.06$ & Ground (CLC) & Levin et al. 1992, Ap.J., 396, 3 \\
9.4   &  3.2  & $3.0  \pm 0.5 $ & Ground (Term)& Roll \& Wilkinson 1966, Phys. Rev. Lett., 16, 405. \\
9.4   &  3.2  & $2.69^{+0.16}_{-0.21}$ & Ground (CLC)& Stokes et al. 1967, Phys. Rev. Lett., 19, 1199. \\
10    &  3.0  & $2.62 \pm 0.06$ & Ground (CLC) & Kogut et al. 1990, Ap.J., 355, 102. \\
10.7  &  2.8  & $2.730\pm 0.014$ & Balloon (LHe) & Staggs et al. 1996, ApJ, 458, 407\\
19.0  &  1.58 & $2.78^{+0.12}_{-0.17}$ & Ground (CLC)& Stokes et al. 1967, Phys. Rev. Lett., 19, 1199. \\
20    &  1.5  & $2.0  \pm 0.4 $ & Ground (CLC) & Welch et al. 1967, Phys. Rev. Lett, 18, 1068. \\
24.8  &  1.2  & $2.783 \pm 0.025$ & Balloon    & Johnson \& Wilkinson 1987, Ap.J. Lett, 313, L1. \\
32.5  &  0.924& $3.16  \pm 0.26$ & Ground (CLC)& Ewing et al. 1967, Phys. Rev. Lett, 19, 1251. \\
33.0  &  0.909& $2.81  \pm 0.12$ & Ground (CLC)& De Amici et al. 1985, Ap.J., 298, 710. \\
35.0  &  0.856& $2.56^{+0.17}_{-0.22}$ & Ground (CLC)& Wilkinson 1967, Phys. Rev. Lett., 19, 1195. \\
37    &  0.82 & $2.9   \pm 0.7 $ & Ground (LN)& Puzanov et al. 1968, Sov. Astr., 11, 905. \\
83.8  & 0.358 & $2.4   \pm 0.7 $ & Ground (LN)& Kislyakov et al. 1971, Sov. Ast., 15, 29. \\
90    &  0.33 & $2.46^{+0.40}_{-0.44}$ & Ground (CLC)&  Boynton et al. 1968, Phys. Rev. Lett., 21, 462. \\
90    &  0.33 & $2.61  \pm 0.25$ & Ground (CLC)& Millea et al. 1971, Phys. Rev. Lett., 26, 919. \\
90    &  0.33 & $2.48  \pm 0.54$ & Plane (Term)& Boynton \& Stokes 1974, Nature, 247, 528. \\
90    &  0.33 & $2.60  \pm 0.09$ & Ground (CLC)& Bersanelli et al. 1989, Ap.J., 339, 632. \\
90.3  &  0.332& $2.97  \pm 1.0 $ &  Balloon  & Bernstein et al. 1990, Ap.J., 362, 107. \\
113.6 & 0.264 & $2.70  \pm 0.04$ & CN (z Per)& Meyer \& Jura 1985, Ap.J., 297, 119. \\
113.6 & 0.264 & $2.74  \pm 0.05$ &  CN (z Oph)& Crane et al. 1986, Ap.J., 309, 12. \\
113.6 & 0.264 & $2.76  \pm 0.07$ & CN (HD 21483)& Meyer et al. 1989, Ap.J. Lett, 343, L1. \\
113.6 & 0.264 & $2.796^{+0.014}_{-0.039}$ & CN (z Oph)& Crane et al. 1989, Ap.J., 346, 136. \\
113.6 & 0.264 & $2.75  \pm 0.04$ & CN (z Per)& Kaiser \& Wright 1990, Ap.J. Lett, 356, L1. \\
113.6 & 0.264 & $2.834 \pm 0.085$ &  CN (HD 154368)& Palazzi et al. 1990, Ap.J., 357, 14. \\
113.6 & 0.264 & $2.807 \pm 0.025$ & CN (16 stars)& Palazzi et al. 1992, Ap.J., 398, 53. \\
154.8 & 0.194 & $3.02  \pm 1.0 $ & Balloon & Bernstein et al. 1990, Ap.J., 362, 107. \\
195.0 & 0.154 & $2.91  \pm 1.0 $ & Balloon & Bernstein et al. 1990, Ap.J., 362, 107. \\
227.3 & 0.132 & $2.76  \pm 0.20$ & CN (z Per)& Meyer \& Jura 1985, Ap.J., 297, 119. \\
227.3 & 0.132 & $2.75^{+0.24}_{-0.29}$ & CN (z Oph)& Crane et al. 1986, Ap.J., 309, 822. \\
227.3 & 0.132 & $2.83  \pm 0.09$ & CN (HD 21483)&  Meyer et al. 1989, Ap.J. Lett, 343, L1. \\
227.3 & 0.132 & $2.832 \pm 0.072$ &  CN (HD 154368)& Palazzi et al. 1990, Ap.J., 357, 14.  \\
266.4 & 0.113 & $2.88  \pm 1.0 $ & Balloon  & Bernstein et al. 1990, Ap.J., 362, 107. \\
\hline
\end{tabular}
\end{center}
\end{table}

\begin{table}[htb]
\begin{center}
\caption{Measurements of $T_{CMB}$}
\begin{tabular}{cccll}
\hspace{-.3 in}Frequency & Wavelength & Temperature & Location & Reference\\
\hspace{-.3 in}(GHz)    &   (cm)     &   (K)       & (calibration) &     \\
\hspace{-.3 in} \tiny	0.408 	& \tiny	  73.5 	& \tiny	 $3.7 \pm 1.2$ 	& \tiny	 Ground (LN)  	& \tiny	 Howell \& Shakeshaft 1967, Nature, 216, 753.   \\
\hspace{-.3 in} \tiny	0.6   	& \tiny	  50   	& \tiny	 $3.0 \pm 1.2$ 	& \tiny	 Ground (Term)	& \tiny	 Sironi et al. 1990, Ap.J., 357, 301.  \\
\hspace{-.3 in} \tiny	0.610 	& \tiny	  49.1 	& \tiny	 $3.7 \pm 1.2$ 	& \tiny	 Ground (LN)  	& \tiny	 Howell \& Shakeshaft 1967, Nature, 216, 7  \\
\hspace{-.3 in} \tiny	0.635 	& \tiny	  47.2 	& \tiny	 $3.0 \pm 0.5$ 	& \tiny	 Ground (LN)  	& \tiny	 Stankevich et al 1970, Australian J. Phys, 23, 529  \\
\hspace{-.3 in} \tiny	0.820 	& \tiny	  36.6 	& \tiny	 $2.7 \pm 1.6$ 	& \tiny	 Ground (Term)	& \tiny	 Sironi et al. 1991, Ap.J., 378, 550.  \\
\hspace{-.3 in} \tiny	1     	& \tiny	  30   	& \tiny	 $2.5 \pm 0.3$ 	& \tiny	 Ground (LN)  	& \tiny	 Pelyushenko \& Stankevich 1969, Sov. Astron., 13, 223.  \\
\hspace{-.3 in} \tiny	1.4   	& \tiny	  21.3 	& \tiny	 $2.11\pm 0.38$ 	& \tiny	 Ground (CLC)	& \tiny	 Levin et al. 1988, Ap.J., 334,14 \\
\hspace{-.3 in} \tiny	1.42  	& \tiny	  21.2 	& \tiny	 $3.2 \pm 1.0$ 	& \tiny	 Ground (Term)	& \tiny	 Penzias and Wilson 1967, AJ, 72, 315  \\
\hspace{-.3 in} \tiny	1.43  	& \tiny	  21 	& \tiny	 $ 2.65^{+0.33}_{-0.30}$ 	& \tiny	 Ground (LN) 	& \tiny	 Staggs et al. 1996, ApJ, 458, 407\\
\hspace{-.3 in} \tiny	1.44  	& \tiny	  20.9 	& \tiny	 $2.5 \pm 0.3$ 	& \tiny	 Ground (LN)  	& \tiny	 Pelyushenko \& Stankevich 1969, Sov. Astron., 13, 223. \\
\hspace{-.3 in} \tiny	1.45  	& \tiny	  20.7 	& \tiny	 $2.8 \pm 0.6$ 	& \tiny	 Ground (Term)	& \tiny	 Howell \& Shakeshaft 1966, Nature, 210, 1318. \\
\hspace{-.3 in} \tiny	1.47  	& \tiny	  20.4 	& \tiny	 $2.27 \pm 0.19$ 	& \tiny	 Ground (CLC)	& \tiny	 Bensadoun et al. 1992 (in press) \\
\hspace{-.3 in} \tiny	2     	& \tiny	  15   	& \tiny	 $2.5 \pm 0.3$ 	& \tiny	 Ground (LN)  	& \tiny	 Pelyushenko \& Stankevich 1969, Sov. Astron., 13, 223. \\
\hspace{-.3 in} \tiny	2.3   	& \tiny	  13.1 	& \tiny	 $2.66 \pm 0.77$ 	& \tiny	 Ground (Term)	& \tiny	 Otoshi \& Stelzreid 1975, IEEE Trans on Inst \& Meas, 24, 174. \\
\hspace{-.3 in} \tiny	2.5   	& \tiny	  12   	& \tiny	 $2.71 \pm 0.21$ 	& \tiny	 Ground (CLC) 	& \tiny	 Sironi et al. 1991, Ap. J., 378, 550. \\
\hspace{-.3 in} \tiny	3.8   	& \tiny	  7.9  	& \tiny	 $2.64 \pm 0.06$ 	& \tiny	 Ground (CLC) 	& \tiny	 De Amici et al. 1991, Ap.J., 381, 341. \\
\hspace{-.3 in} \tiny	4.08  	& \tiny	  7.35 	& \tiny	 $3.5  \pm 1.0 $ 	& \tiny	 Ground (Term)	& \tiny	 Penzias \& Wilson 1965, Ap.J., 142, 419. \\
\hspace{-.3 in} \tiny	4.75  	& \tiny	  6.3  	& \tiny	 $2.70 \pm 0.07$ 	& \tiny	 Ground (CLC) 	& \tiny	 Mandolesi et al. 1986, Ap.J., 310, 561. \\
\hspace{-.3 in} \tiny	7.5   	& \tiny	  4.0  	& \tiny	 $2.60 \pm 0.07$ 	& \tiny	 Ground (CLC) 	& \tiny	 Kogut et al. 1988, Ap.J., 355, 102\\
\hspace{-.3 in} \tiny	7.5   	& \tiny	  4.0  	& \tiny	 $2.64 \pm 0.06$ 	& \tiny	 Ground (CLC) 	& \tiny	 Levin et al. 1992, Ap.J., 396, 3 \\
\hspace{-.3 in} \tiny	9.4   	& \tiny	  3.2  	& \tiny	 $3.0  \pm 0.5 $ 	& \tiny	 Ground (Term)	& \tiny	 Roll \& Wilkinson 1966, Phys. Rev. Lett., 16, 405. \\
\hspace{-.3 in} \tiny	9.4   	& \tiny	  3.2  	& \tiny	 $2.69^{+0.16}_{-0.21}$ 	& \tiny	 Ground (CLC)	& \tiny	 Stokes et al. 1967, Phys. Rev. Lett., 19, 1199. \\
\hspace{-.3 in} \tiny	10    	& \tiny	  3.0  	& \tiny	 $2.62 \pm 0.06$ 	& \tiny	 Ground (CLC) 	& \tiny	 Kogut et al. 1990, Ap.J., 355, 102. \\
\hspace{-.3 in} \tiny	10.7  	& \tiny	  2.8  	& \tiny	 $2.730\pm 0.014$ 	& \tiny	 Balloon (LHe) 	& \tiny	 Staggs et al. 1996, ApJ, 458, 407\\
\hspace{-.3 in} \tiny	19.0  	& \tiny	  1.58 	& \tiny	 $2.78^{+0.12}_{-0.17}$ 	& \tiny	 Ground (CLC)	& \tiny	 Stokes et al. 1967, Phys. Rev. Lett., 19, 1199. \\
\hspace{-.3 in} \tiny	20    	& \tiny	  1.5  	& \tiny	 $2.0  \pm 0.4 $ 	& \tiny	 Ground (CLC) 	& \tiny	 Welch et al. 1967, Phys. Rev. Lett, 18, 1068. \\
\hspace{-.3 in} \tiny	24.8  	& \tiny	  1.2  	& \tiny	 $2.783 \pm 0.025$ 	& \tiny	 Balloon    	& \tiny	 Johnson \& Wilkinson 1987, Ap.J. Lett, 313, L1. \\
\end{tabular}
\end{center}
\end{table}



\begin{table}[htb]
\begin{center}
\caption{Measurements of $T_{CMB}$ Continued}
\begin{tabular}{cccll}
\hspace{-.3 in}Frequency & Wavelength & Temperature & Location & Reference\\
\hspace{-.3 in}(GHz)    &   (cm)     &   (K)       & (calibration) &     \\
\hspace{-.3 in} \tiny	32.5  	& \tiny	  0.924	& \tiny	 $3.16  \pm 0.26$ 	& \tiny	 Ground (CLC)	& \tiny	 Ewing et al. 1967, Phys. Rev. Lett, 19, 1251. \\
\hspace{-.3 in} \tiny	33.0  	& \tiny	  0.909	& \tiny	 $2.81  \pm 0.12$ 	& \tiny	 Ground (CLC)	& \tiny	 De Amici et al. 1985, Ap.J., 298, 710. \\
\hspace{-.3 in} \tiny	35.0  	& \tiny	  0.856	& \tiny	 $2.56^{+0.17}_{-0.22}$ 	& \tiny	 Ground (CLC)	& \tiny	 Wilkinson 1967, Phys. Rev. Lett., 19, 1195. \\
\hspace{-.3 in} \tiny	37    	& \tiny	  0.82 	& \tiny	 $2.9   \pm 0.7 $ 	& \tiny	 Ground (LN)	& \tiny	 Puzanov et al. 1968, Sov. Astr., 11, 905. \\
\hspace{-.3 in} \tiny	83.8  	& \tiny	 0.358 	& \tiny	 $2.4   \pm 0.7 $ 	& \tiny	 Ground (LN)	& \tiny	 Kislyakov et al. 1971, Sov. Ast., 15, 29. \\
\hspace{-.3 in} \tiny	90    	& \tiny	  0.33 	& \tiny	 $2.46^{+0.40}_{-0.44}$ 	& \tiny	 Ground (CLC)	& \tiny	  Boynton et al. 1968, Phys. Rev. Lett., 21, 462. \\
\hspace{-.3 in} \tiny	90    	& \tiny	  0.33 	& \tiny	 $2.61  \pm 0.25$ 	& \tiny	 Ground (CLC)	& \tiny	 Millea et al. 1971, Phys. Rev. Lett., 26, 919. \\
\hspace{-.3 in} \tiny	90    	& \tiny	  0.33 	& \tiny	 $2.48  \pm 0.54$ 	& \tiny	 Plane (Term)	& \tiny	 Boynton \& Stokes 1974, Nature, 247, 528. \\
\hspace{-.3 in} \tiny	90    	& \tiny	  0.33 	& \tiny	 $2.60  \pm 0.09$ 	& \tiny	 Ground (CLC)	& \tiny	 Bersanelli et al. 1989, Ap.J., 339, 632. \\
\hspace{-.3 in} \tiny	90.3  	& \tiny	  0.332	& \tiny	 $2.97  \pm 1.0 $ 	& \tiny	  Balloon  	& \tiny	 Bernstein et al. 1990, Ap.J., 362, 107. \\
\hspace{-.3 in} \tiny	113.6 	& \tiny	 0.264 	& \tiny	 $2.70  \pm 0.04$ 	& \tiny	 CN (z Per)	& \tiny	 Meyer \& Jura 1985, Ap.J., 297, 119. \\
\hspace{-.3 in} \tiny	113.6 	& \tiny	 0.264 	& \tiny	 $2.74  \pm 0.05$ 	& \tiny	  CN (z Oph)	& \tiny	 Crane et al. 1986, Ap.J., 309, 12. \\
\hspace{-.3 in} \tiny	113.6 	& \tiny	 0.264 	& \tiny	 $2.76  \pm 0.07$ 	& \tiny	 CN (HD 21483)	& \tiny	 Meyer et al. 1989, Ap.J. Lett, 343, L1. \\
\hspace{-.3 in} \tiny	113.6 	& \tiny	 0.264 	& \tiny	 $2.796^{+0.014}_{-0.039}$ 	& \tiny	 CN (z Oph)	& \tiny	 Crane et al. 1989, Ap.J., 346, 136. \\
\hspace{-.3 in} \tiny	113.6 	& \tiny	 0.264 	& \tiny	 $2.75  \pm 0.04$ 	& \tiny	 CN (z Per)	& \tiny	 Kaiser \& Wright 1990, Ap.J. Lett, 356, L1. \\
\hspace{-.3 in} \tiny	113.6 	& \tiny	 0.264 	& \tiny	 $2.834 \pm 0.085$ 	& \tiny	  CN (HD 154368)	& \tiny	 Palazzi et al. 1990, Ap.J., 357, 14. \\
\hspace{-.3 in} \tiny	113.6 	& \tiny	 0.264 	& \tiny	 $2.807 \pm 0.025$ 	& \tiny	 CN (16 stars)	& \tiny	 Palazzi et al. 1992, Ap.J., 398, 53. \\
\hspace{-.3 in} \tiny	154.8 	& \tiny	 0.194 	& \tiny	 $3.02  \pm 1.0 $ 	& \tiny	 Balloon 	& \tiny	 Bernstein et al. 1990, Ap.J., 362, 107. \\
\hspace{-.3 in} \tiny	195.0 	& \tiny	 0.154 	& \tiny	 $2.91  \pm 1.0 $ 	& \tiny	 Balloon 	& \tiny	 Bernstein et al. 1990, Ap.J., 362, 107. \\
\hspace{-.3 in} \tiny	227.3 	& \tiny	 0.132 	& \tiny	 $2.76  \pm 0.20$ 	& \tiny	 CN (z Per)	& \tiny	 Meyer \& Jura 1985, Ap.J., 297, 119. \\
\hspace{-.3 in} \tiny	227.3 	& \tiny	 0.132 	& \tiny	 $2.75^{+0.24}_{-0.29}$ 	& \tiny	 CN (z Oph)	& \tiny	 Crane et al. 1986, Ap.J., 309, 822. \\
\hspace{-.3 in} \tiny	227.3 	& \tiny	 0.132 	& \tiny	 $2.83  \pm 0.09$ 	& \tiny	 CN (HD 21483)	& \tiny	  Meyer et al. 1989, Ap.J. Lett, 343, L1. \\
\hspace{-.3 in} \tiny	227.3 	& \tiny	 0.132 	& \tiny	 $2.832 \pm 0.072$ 	& \tiny	  CN (HD 154368)	& \tiny	 Palazzi et al. 1990, Ap.J., 357, 14.  \\
\hspace{-.3 in} \tiny	266.4 	& \tiny	 0.113 	& \tiny	 $2.88  \pm 1.0 $ 	& \tiny	 Balloon  	& \tiny	 Bernstein et al. 1990, Ap.J., 362, 107. \\
\end{tabular}
\end{center}
\end{table}


\clearpage
\begin{table*}
\caption{FIRAS Data;\ \protect{\cite{fixsen96}}}
\begin{center}
\begin{tabular}{rrrr}
Frequency & Brightness ($B_\nu$) & Upper Error & Lower Error\\
GHz & erg/(cm$^{2}$ s sr Hz) & erg/(cm$^{2}$ s sr Hz) & erg/(cm$^{2}$ s sr Hz) \\
$68.1$&$20134.\times10^{-19}$&$14.\times10^{-19}$&$14.\times10^{-19}$\\
$81.6$&$25039.\times10^{-19}$&$19.\times10^{-19}$&$19.\times10^{-19}$\\
$95.4$&$29424.\times10^{-19}$&$25.\times10^{-19}$&$25.\times10^{-19}$\\
$108.9$&$32924.\times10^{-19}$&$23.\times10^{-19}$&$23.\times10^{-19}$\\
$122.4$&$35591.\times10^{-19}$&$22.\times10^{-19}$&$22.\times10^{-19}$\\
$136.2$&$37385.\times10^{-19}$&$21.\times10^{-19}$&$21.\times10^{-19}$\\
$149.7$&$38308.\times10^{-19}$&$18.\times10^{-19}$&$18.\times10^{-19}$\\
$163.5$&$38500.\times10^{-19}$&$18.\times10^{-19}$&$18.\times10^{-19}$\\
$177.0$&$38073.\times10^{-19}$&$16.\times10^{-19}$&$16.\times10^{-19}$\\
$190.5$&$37074.\times10^{-19}$&$14.\times10^{-19}$&$14.\times10^{-19}$\\
$204.3$&$35595.\times10^{-19}$&$13.\times10^{-19}$&$13.\times10^{-19}$\\
$217.8$&$33825.\times10^{-19}$&$12.\times10^{-19}$&$12.\times10^{-19}$\\
$231.3$&$31781.\times10^{-19}$&$11.\times10^{-19}$&$11.\times10^{-19}$\\
$245.1$&$29564.\times10^{-19}$&$10.\times10^{-19}$&$10.\times10^{-19}$\\
$258.6$&$27314.\times10^{-19}$&$11.\times10^{-19}$&$11.\times10^{-19}$\\
$272.4$&$24968.\times10^{-19}$&$12.\times10^{-19}$&$12.\times10^{-19}$\\
$285.9$&$22735.\times10^{-19}$&$14.\times10^{-19}$&$14.\times10^{-19}$\\
$299.4$&$20575.\times10^{-19}$&$16.\times10^{-19}$&$16.\times10^{-19}$\\
$313.2$&$18469.\times10^{-19}$&$18.\times10^{-19}$&$18.\times10^{-19}$\\
$326.7$&$16489.\times10^{-19}$&$22.\times10^{-19}$&$22.\times10^{-19}$\\
$340.2$&$14628.\times10^{-19}$&$22.\times10^{-19}$&$22.\times10^{-19}$\\
$354.0$&$12979.\times10^{-19}$&$23.\times10^{-19}$&$23.\times10^{-19}$\\
$367.5$&$11454.\times10^{-19}$&$23.\times10^{-19}$&$23.\times10^{-19}$\\
$381.3$&$10028.\times10^{-19}$&$23.\times10^{-19}$&$23.\times10^{-19}$\\
$394.8$&$8783.\times10^{-19}$&$22.\times10^{-19}$&$22.\times10^{-19}$\\
$408.3$&$7644.\times10^{-19}$&$21.\times10^{-19}$&$21.\times10^{-19}$\\
$422.1$&$6630.\times10^{-19}$&$20.\times10^{-19}$&$20.\times10^{-19}$\\
$435.6$&$5772.\times10^{-19}$&$19.\times10^{-19}$&$19.\times10^{-19}$\\

\end{tabular}
\end{center}
\label{firasb}
\end{table*}

\begin{table*}
\caption{FIRAS Data -- continued;\ \protect{\cite{fixsen96}}}
\begin{center}
\begin{tabular}{rrrr}
Frequency & Brightness ($B_\nu$) & Upper Error & Lower Error\\
GHz & erg/(cm$^{2}$ s sr Hz) & erg/(cm$^{2}$ s sr Hz) & erg/(cm$^{2}$ s sr Hz) \\
$449.1$&$4977.\times10^{-19}$&$19.\times10^{-19}$&$19.\times10^{-19}$\\
$462.9$&$4257.\times10^{-19}$&$19.\times10^{-19}$&$19.\times10^{-19}$\\
$476.4$&$3667.\times10^{-19}$&$21.\times10^{-19}$&$21.\times10^{-19}$\\
$490.2$&$3141.\times10^{-19}$&$23.\times10^{-19}$&$23.\times10^{-19}$\\
$503.7$&$2702.\times10^{-19}$&$26.\times10^{-19}$&$26.\times10^{-19}$\\
$517.2$&$2274.\times10^{-19}$&$28.\times10^{-19}$&$28.\times10^{-19}$\\
$531.0$&$1936.\times10^{-19}$&$30.\times10^{-19}$&$30.\times10^{-19}$\\
$544.5$&$1672.\times10^{-19}$&$32.\times10^{-19}$&$32.\times10^{-19}$\\
$558.3$&$1389.\times10^{-19}$&$33.\times10^{-19}$&$33.\times10^{-19}$\\
$571.8$&$1170.\times10^{-19}$&$35.\times10^{-19}$&$35.\times10^{-19}$\\
$585.3$&$998.\times10^{-19}$&$41.\times10^{-19}$&$41.\times10^{-19}$\\
$599.1$&$865.\times10^{-19}$&$55.\times10^{-19}$&$55.\times10^{-19}$\\
$612.6$&$787.\times10^{-19}$&$88.\times10^{-19}$&$88.\times10^{-19}$\\
$626.1$&$548.\times10^{-19}$&$155.\times10^{-19}$&$155.\times10^{-19}$\\
$639.9$&$26.\times10^{-19}$&$282.\times10^{-19}$&$282.\times10^{-19}$

\end{tabular}
\end{center}
\end{table*}

\begin{table*}
\caption{FIRAS Data;\ \protect{\cite{fixsen96}}}
\begin{center}
\begin{tabular}{rrrr}
Frequency & Temperature & Upper Error & Lower Error\\
GHz & K & K & K\\
$68.1$&$2.72839$&$0.00111$&$0.00111$\\
$81.6$&$2.72892$&$0.00110$&$0.00110$\\
$95.4$&$2.72959$&$0.00112$&$0.00112$\\
$108.9$&$2.72974$&$0.00085$&$0.00085$\\
$122.4$&$2.73034$&$0.00069$&$0.00069$\\
$136.2$&$2.72951$&$0.00058$&$0.00058$\\
$149.7$&$2.72875$&$0.00045$&$0.00045$\\
$163.5$&$2.72852$&$0.00042$&$0.00042$\\
$177.0$&$2.72922$&$0.00035$&$0.00035$\\
$190.5$&$2.72931$&$0.00030$&$0.00030$\\
$204.3$&$2.72927$&$0.00027$&$0.00027$\\
$217.8$&$2.72954$&$0.00025$&$0.00025$\\
$231.3$&$2.72908$&$0.00023$&$0.00023$\\
$245.1$&$2.72925$&$0.00021$&$0.00021$\\
$258.6$&$2.72942$&$0.00024$&$0.00024$\\
$272.4$&$2.72895$&$0.00027$&$0.00027$\\
$285.9$&$2.72916$&$0.00033$&$0.00033$\\
$299.4$&$2.72946$&$0.00040$&$0.00040$\\
$313.2$&$2.72976$&$0.00048$&$0.00048$\\
$326.7$&$2.72892$&$0.00063$&$0.00063$\\
$340.2$&$2.72750$&$0.00068$&$0.00068$\\
$354.0$&$2.72945$&$0.00078$&$0.00078$\\
$367.5$&$2.72968$&$0.00085$&$0.00085$\\

\end{tabular}
\end{center}
\label{firasb}
\end{table*}

\begin{table*}
\caption{FIRAS Data -- continued;\ \protect{\cite{fixsen96}}}
\begin{center}
\begin{tabular}{rrrr}
Frequency & Temperature & Upper Error & Lower Error\\
GHz & K & K & K\\
$394.8$&$2.72970$&$0.00098$&$0.00098$\\
$408.3$&$2.72886$&$0.00104$&$0.00104$\\
$422.1$&$2.72919$&$0.00111$&$0.00111$\\
$435.6$&$2.73080$&$0.00117$&$0.00118$\\
$449.1$&$2.72997$&$0.00132$&$0.00132$\\
$462.9$&$2.72847$&$0.00149$&$0.00150$\\
$476.4$&$2.72918$&$0.00186$&$0.00187$\\
$490.2$&$2.72989$&$0.00231$&$0.00233$\\
$503.7$&$2.73149$&$0.00296$&$0.00298$\\
$517.2$&$2.72717$&$0.00367$&$0.00371$\\
$531.0$&$2.72808$&$0.00450$&$0.00455$\\
$544.5$&$2.73254$&$0.00543$&$0.00551$\\
$558.3$&$2.72739$&$0.00653$&$0.00666$\\
$571.8$&$2.72600$&$0.00800$&$0.00820$\\
$585.3$&$2.72813$&$0.01071$&$0.01107$\\
$599.1$&$2.73605$&$0.01614$&$0.01699$\\
$612.6$&$2.75533$&$0.02764$&$0.03028$\\
$626.1$&$2.70767$&$0.06215$&$0.07872$\\
$639.9$&$2.16250$&$0.45372$&$1.16250$\\

\end{tabular}
\end{center}
\end{table*}

\begin{table*}
\caption{UBC COBRA Rocket Data;\ \protect{\cite{gush90}}}
\begin{center}
\begin{tabular}{rrrr}
Frequency & Brightness ($B_\nu$) & Upper Error & Lower Error\\
GHz & erg/(cm$^{2}$ s sr Hz) & erg/(cm$^{2}$ s sr Hz) & erg/(cm$^{2}$ s sr Hz) \\
$54.6$&$1.5398\times10^{-15}$&$1.4482\times10^{-17}$&$1.4469\times10^{-17}$\\
$68.1$&$1.9120\times10^{-15}$&$2.1388\times10^{-17}$&$2.1354\times10^{-17}$\\
$81.9$&$2.4017\times10^{-15}$&$2.9351\times10^{-17}$&$2.9285\times10^{-17}$\\
$95.4$&$2.9589\times10^{-15}$&$3.7901\times10^{-17}$&$3.7796\times10^{-17}$\\
$109.2$&$3.2674\times10^{-15}$&$4.6194\times10^{-17}$&$4.6026\times10^{-17}$\\
$122.7$&$3.5439\times10^{-15}$&$5.4155\times10^{-17}$&$5.3910\times10^{-17}$\\
$136.5$&$3.8268\times10^{-15}$&$6.2135\times10^{-17}$&$6.1804\times10^{-17}$\\
$150.0$&$3.8609\times10^{-15}$&$6.8239\times10^{-17}$&$6.7800\times10^{-17}$\\
$163.8$&$3.8302\times10^{-15}$&$7.3341\times10^{-17}$&$7.2782\times10^{-17}$\\
$177.3$&$3.8319\times10^{-15}$&$7.7897\times10^{-17}$&$7.7221\times10^{-17}$\\
$191.1$&$3.7099\times10^{-15}$&$8.0751\times10^{-17}$&$7.9950\times10^{-17}$\\
$204.6$&$3.5353\times10^{-15}$&$8.2160\times10^{-17}$&$8.1240\times10^{-17}$\\
$218.4$&$3.3956\times10^{-15}$&$8.3253\times10^{-17}$&$8.2222\times10^{-17}$\\
$231.9$&$3.2069\times10^{-15}$&$8.2976\times10^{-17}$&$8.1846\times10^{-17}$\\
$245.7$&$2.9737\times10^{-15}$&$8.1387\times10^{-17}$&$8.0168\times10^{-17}$\\
$259.2$&$2.7378\times10^{-15}$&$7.8988\times10^{-17}$&$7.7697\times10^{-17}$\\
$273.0$&$2.5046\times10^{-15}$&$7.5987\times10^{-17}$&$7.4640\times10^{-17}$\\
$286.5$&$2.2883\times10^{-15}$&$7.2673\times10^{-17}$&$7.1289\times10^{-17}$\\
$300.3$&$2.0654\times10^{-15}$&$6.8708\times10^{-17}$&$6.7302\times10^{-17}$\\
$586.8$&$9.6038\times10^{-17}$&$6.3650\times10^{-18}$&$6.0420\times10^{-18}$\\

\end{tabular}
\end{center}
\label{ubcb}
\end{table*}

\begin{table*}
\caption{UBC COBRA Rocket Data -- continued;\ \protect{\cite{gush90}}}
\begin{center}
\begin{tabular}{rrrr}
Frequency & Brightness ($B_\nu$) & Upper Error & Lower Error\\
GHz & erg/(cm$^{2}$ s sr Hz) & erg/(cm$^{2}$ s sr Hz) & erg/(cm$^{2}$ s sr Hz) \\
$327.6$&$1.7248\times10^{-15}$&$6.1696\times10^{-17}$&$6.0285\times10^{-17}$\\
$341.1$&$1.5014\times10^{-15}$&$5.6371\times10^{-17}$&$5.4991\times10^{-17}$\\
$354.9$&$1.3007\times10^{-15}$&$5.1158\times10^{-17}$&$4.9822\times10^{-17}$\\
$368.4$&$1.1943\times10^{-15}$&$4.8165\times10^{-17}$&$4.6860\times10^{-17}$\\
$382.2$&$1.0504\times10^{-15}$&$4.3935\times10^{-17}$&$4.2682\times10^{-17}$\\
$395.7$&$8.9122\times10^{-16}$&$3.8983\times10^{-17}$&$3.7804\times10^{-17}$\\
$409.5$&$8.0218\times10^{-16}$&$3.5984\times10^{-17}$&$3.4856\times10^{-17}$\\
$423.0$&$7.3678\times10^{-16}$&$3.3659\times10^{-17}$&$3.2576\times10^{-17}$\\
$436.8$&$5.9625\times10^{-16}$&$2.8669\times10^{-17}$&$2.7686\times10^{-17}$\\
$450.3$&$5.2788\times10^{-16}$&$2.6027\times10^{-17}$&$2.5104\times10^{-17}$\\
$464.1$&$4.6379\times10^{-16}$&$2.3462\times10^{-17}$&$2.2601\times10^{-17}$\\
$477.6$&$3.1606\times10^{-16}$&$1.7431\times10^{-17}$&$1.6724\times10^{-17}$\\
$491.4$&$2.2851\times10^{-16}$&$1.3475\times10^{-17}$&$1.2885\times10^{-17}$\\
$504.9$&$2.2695\times10^{-16}$&$1.3282\times10^{-17}$&$1.2703\times10^{-17}$\\
$518.7$&$2.2705\times10^{-16}$&$1.3163\times10^{-17}$&$1.2593\times10^{-17}$\\
$532.2$&$1.8120\times10^{-16}$&$1.0942\times10^{-17}$&$1.0445\times10^{-17}$\\
$546.0$&$1.5658\times10^{-16}$&$9.6635\times10^{-18}$&$9.2128\times10^{-18}$\\
$559.5$&$1.4474\times10^{-16}$&$8.9965\times10^{-18}$&$8.5725\times10^{-18}$\\
$573.3$&$1.1111\times10^{-16}$&$7.2222\times10^{-18}$&$6.8641\times10^{-18}$\\
$586.8$&$9.6038\times10^{-17}$&$6.3650\times10^{-18}$&$6.0420\times10^{-18}$\\

\end{tabular}
\end{center}
\label{ubcb2}
\end{table*}

\begin{table*}
\caption{UBC COBRA Rocket Data;\ \protect{\cite{gush90}}}
\begin{center}
\begin{tabular}{rrrr}
Frequency & Temperature & Upper Error & Lower Error\\
GHz & K & K & K\\
$54.6$&$2.789$&$0.017$&$0.017$\\
$68.1$&$2.648$&$0.017$&$0.017$\\
$81.9$&$2.664$&$0.017$&$0.017$\\
$95.4$&$2.737$&$0.017$&$0.017$\\
$109.2$&$2.718$&$0.017$&$0.017$\\
$122.7$&$2.724$&$0.017$&$0.017$\\
$136.5$&$2.753$&$0.017$&$0.017$\\
$150.0$&$2.736$&$0.017$&$0.017$\\
$163.8$&$2.724$&$0.017$&$0.017$\\
$177.3$&$2.735$&$0.017$&$0.017$\\
$191.1$&$2.731$&$0.017$&$0.017$\\
$204.6$&$2.725$&$0.017$&$0.017$\\
$218.4$&$2.734$&$0.017$&$0.017$\\
$231.9$&$2.737$&$0.017$&$0.017$\\
$245.7$&$2.735$&$0.017$&$0.017$\\
$259.2$&$2.733$&$0.017$&$0.017$\\
$273.0$&$2.733$&$0.017$&$0.017$\\
$286.5$&$2.735$&$0.017$&$0.017$\\
$300.3$&$2.735$&$0.017$&$0.017$\\

\end{tabular}
\end{center}
\label{ubct}
\end{table*}

\begin{table*}
\caption{UBC COBRA Rocket Data -- continued;\ \protect{\cite{gush90}}}
\begin{center}
\begin{tabular}{rrrr}
Frequency & Temperature & Upper Error & Lower Error\\
GHz & K & K & K\\
$313.8$&$2.742$&$0.017$&$0.017$\\
$327.6$&$2.754$&$0.017$&$0.017$\\
$341.1$&$2.743$&$0.017$&$0.017$\\
$354.9$&$2.734$&$0.017$&$0.017$\\
$368.4$&$2.751$&$0.017$&$0.017$\\
$382.2$&$2.752$&$0.017$&$0.017$\\
$395.7$&$2.739$&$0.017$&$0.017$\\
$409.5$&$2.752$&$0.017$&$0.017$\\
$423.0$&$2.772$&$0.017$&$0.017$\\
$436.8$&$2.747$&$0.017$&$0.017$\\
$450.3$&$2.755$&$0.017$&$0.017$\\
$464.1$&$2.762$&$0.017$&$0.017$\\
$477.6$&$2.686$&$0.017$&$0.017$\\
$491.4$&$2.637$&$0.017$&$0.017$\\
$504.9$&$2.683$&$0.017$&$0.017$\\
$518.7$&$2.732$&$0.017$&$0.017$\\
$532.2$&$2.713$&$0.017$&$0.017$\\
$546.0$&$2.719$&$0.017$&$0.017$\\
$559.5$&$2.743$&$0.017$&$0.017$\\
$573.3$&$2.717$&$0.017$&$0.017$\\
$586.8$&$2.723$&$0.017$&$0.017$\\

\end{tabular}
\end{center}
\label{ubct2}
\end{table*}

\end{document}